\PassOptionsToPackage{svgnames, prologue, dvipsnames}{xcolor}

\documentclass[a4paper]{article}

\usepackage{xcolor}
\usepackage{hyperref}
\usepackage{color}

\hypersetup{
  pdfborder=0 0 0,
}

\usepackage{times}

\usepackage{booktabs}       

\usepackage{csquotes}
\usepackage{amsmath}
\usepackage{amsfonts}
\usepackage{bm}

\usepackage{cleveref}
\usepackage{fontawesome}
\usepackage[nohypertypes={acronym}]{glossaries}
\usepackage{xspace}
\usepackage{subcaption}
\usepackage{multirow}
\usepackage{mathtools} 
\usepackage{wrapfig}

\usepackage{rotating}
\usepackage[natbib, style=numeric-comp, sorting=none, sortcites, maxbibnames=15]{biblatex}

\usepackage{tabularray}
\usepackage{pgfplots}    
\usepackage{authblk} 
\usepackage{comment} 
\usepackage{geometry}
\usepackage{multicol}
\usepackage{dcolumn}
\usetikzlibrary{external}
\usepackage{fancyhdr}
\glsdisablehyper
\usetikzlibrary{shapes,shapes.misc, graphs, patterns,shapes.arrows, arrows.meta, positioning, fit, arrows.meta, graphs, shapes.misc, matrix, quotes, patterns, shapes.arrows, shapes.misc, positioning, calc, fit, bending}
\usepgfplotslibrary{groupplots,dateplot}
\pgfplotsset{compat=1.18}
\DeclareUnicodeCharacter{2212}{−}

\graphicspath{{figures/}{pictures/}{images/}{./}} 

\newcommand{\speedupBsbmBestScanEmbeddedNeoj}{$2.61 \times  $\xspace}

\newcommand{\speedupBsbmBestScanEmbeddedSystemnamebase}{$3.15 \times  $\xspace}

\newcommand{\speedupBsbmWorstScanEmbeddedSystemnamebase}{$1.56 \times  $\xspace}

\newcommand{\bestSpeedupBsbmScanEmbedded}{$41.9 \times  $\xspace}

\newcommand{\worstSpeedupBsbmScanEmbedded}{$3.42 \times  $\xspace}

\newcommand{\speedupBsbmBestScanIndexSystemnamebase}{$355 \times  $\xspace}

\newcommand{\bestSpeedupBsbmScanIndex}{$250 \times  $\xspace}

\newcommand{\worstSpeedupBsbmScanIndex}{$3.03 \times  $\xspace}

\newcommand{\bestSpeedupBsbmJoinEmbedded}{$25.8 \times  $\xspace}

\newcommand{\worstSpeedupBsbmJoinEmbedded}{$4.11 \times  $\xspace}

\newcommand{\bestSpeedupBsbmJoinIndex}{$188 \times  $\xspace}

\newcommand{\bestSpeedupDbpediaScanIndex}{$97.8 \times  $\xspace}

\newcommand{\speedupDbpediaBestScanEmbeddedSystemnamebase}{$1.52 \times  $\xspace}

\newcommand{\speedupDbpediaBestScanEmbeddedSystemnametv}{$20 \times  $\xspace}

\newcommand{\bestSpeedupDbpediaScanEmbedded}{$20 \times  $\xspace}

\newacronym[shortplural=LMs, longplural=Language Models]{lm}{LM}{Language Model}
\newacronym[shortplural=KGs, longplural=Knowledge Graphs]{kg}{KG}{Knowledge Graph}
\newacronym[shortplural=GUIs, longplural=Graphical User Interfaces]{gui}{GUI}{Graphical User Interface}
\newacronym{ged}{GED}{Graph Edit Distance}
\newacronym{rag}{RAG}{Retrieval Augmented Generation}
\newacronym{qa}{QA}{Question-Answering}
\newacronym{oql}{OQL}{Ontology Query Language}
\newacronym[shortplural=BGPs, longplural=Basic Graph Patterns]{bgp}{BGP}{Basic Graph Pattern}
\newacronym{gbnf}{GBNF}{GGML Backus-Naur Form}
\newacronym{onset}{OnSET}{Ontology and Semantic Exploration Toolkit}
\newacronym{bto}{BTO}{Brainteaser Ontology}
\newacronym{als}{ALS}{Amyotrophic lateral sclerosis}
\newacronym{ms}{MS}{Multiple sclerosis}
\newacronym{ir}{IR}{Information Retrival}
\newacronym{bsbm}{BSBM}{Berlin Sparql Benchmark}
\newacronym{json}{JSON}{Javascript Object Notation}
\newacronym[shortplural=URIs, longplural=Unique Resource Identifiers]{uri}{URI}{Unique Resource Identifier}
\newacronym[shortplural=URLs, longplural=Unique Resource Links]{url}{URL}{Unique Resource Link}
\newacronym[shortplural=kNNs, longplural=$k$-Nearest Neighbours]{knn}{kNN}{$k$-Nearest Neighbours}
\newacronym[shortplural=AkNNs, longplural=Approximate $k$-Nearest Neighbours]{aknn}{AkNN}{Approximate $k$-Nearest Neighbours}
\newacronym{wit}{WIT}{Wikipedia-based Image Text}
\newacronym[shortplural=DBMSs, longplural=Database Management Systems]{dbms}{DBMS}{Database Management System}
\newacronym{rdf}{RDF}{Resource Description Framework}
\newacronym[shortplural=PGs, longplural=Property Graphs]{pg}{PG}{Property Graph}
\newacronym{hnsw}{HNSW}{Hierarchical Navigable Small World}
\newacronym{ivf}{IVF}{Inverted File}
\newacronym[shortplural=VDBMSs, longplural=Vector Database Management Systems]{vdbms}{VDBMS}{Vector Database Management System}
\newacronym{dvq}{DVQ}{Dense Vector Querying}
\newacronym{bsbmdvq}{\acrshort{bsbm}-\acrshort{dvq}}{\acrlong{bsbm} for \acrlong{dvq}}
\newacronym{mus}{MS}{Multimodal Search}
\newacronym{dbpediams}{DBpedia-\acrshort{ms}}{DBpedia-\acrlong{ms}}
\newacronym{fvs}{FVS}{Filtered Vector Search}
\newacronym{tenor}{TENOR}{TENsor Operators for RDF}
\newacronym{everoc}{EVeRoK}{Efficient Vector Retrieval over Knowledge Graphs}
\newacronym{quiver}{QUIVER}{QLever-Unified Indexed Vector Embedding Retrieval }

\newacronym[shortplural=IAIs, longplural=interpretable artifical intelligence]{iai}{IAI}{interpretable artificial intelligence}
\newacronym[shortplural=DNNs, longplural=deep neural networks]{dnn}{DNN}{deep neural network}
\newacronym[shortplural=NNs, longplural=neural networks]{nn}{NN}{neural network}
\newacronym[shortplural=EAFs, longplural=electric arc furnaces]{eaf}{EAF}{electric arc furnace}
\newacronym[shortplural=ALEs, longplural=averaged local effects]{ale}{ALE}{averaged local effects}
\newacronym[shortplural=PDPs, longplural=partial dependence plots]{pdp}{PDP}{partial dependence plot}
\newacronym{ml}{ML}{Machine Learning}
\newacronym{ai}{AI}{Artificial Intelligence}
\newacronym{lime}{LIME}{local interpretable model-agnostic explanations}
\newacronym{shap}{SHAP}{Shapley additive explanations}
\newacronym{som}{SOM}{Self-Organizing Maps}
\newacronym{iid}{iid.}{independent and identically distributed}
\newacronym{mse}{MSE}{mean squared error}
\newacronym{mae}{MAE}{mean absolute error}
\newacronym{relu}{ReLU}{rectified linear unit}
\newacronym{xai}{XAI}{eXplainable Artificial Intelligence}
\newacronym{sg}{SG}{SmoothGrad}
\newacronym{ig}{IG}{integrated gradients}
\newacronym{ofc}{OFC}{optical flow constraint}
\newacronym[shortplural=GMMs, longplural=Gaussian mixture models]{gmm}{GMM}{Gaussian mixture model}
\newacronym[shortplural=GPs, longplural=Gaussian processes]{gp}{GP}{Gaussian process}
\newacronym{ood}{OOD}{out-of-domain}
\newacronym{ucq}{UCQ}{Uncertainty Quantification}
\newacronym[shortplural=LGBMs, longplural=Light Gradient Boosting Machines]{lgbm}{LightGBM}{Light Gradient Boosting Machine}
\newacronym{map}{MAP}{maximum a posteriori}
\newacronym{kl}{KL}{Kullback-Leibler}
\newacronym{mc}{MC}{monte-carlo}
\newacronym{gda}{GDA}{gaussian discriminatory analysis}
\newacronym{ddu}{DDU}{deep deterministic uncertainty}
\newacronym[shortplural=DEs, longplural=deep ensembles]{de}{DE}{deep ensemble}
\newacronym{clue}{CLUE}{counterfactual latent uncertainty explanations}
\newacronym{vae}{VAE}{Variational Autoencoder}
\newacronym{rbf}{RBF}{radial basis function}
\newacronym{se}{SE}{squared exponential}
\newacronym{tsne}{t-SNE}{t-Distributed Stochastic Neighbor Embedding}
\newacronym[shortplural=SCAs, longplural=smoothness constrained attributions]{sca}{SCA}{smoothness constrained attribution}
\newacronym[shortplural=PIs, longplural=prediction intervals]{pi}{PI}{prediction interval}
\newacronym[shortplural=PIOs, longplural=prediction interval overlap]{pio}{PIO}{prediction interval overlap}
\newacronym[shortplural=PICPs, longplural=prediction interval coverage probability]{picp}{PICP}{prediction interval coverage probability}
\newacronym[shortplural=iUCAMs, longplural=input uncertainty attribution mechanisms]{iuam}{iUCAM}{input uncertainty attribution mechanism}
\newacronym{mast}{MAST}{Multi-Agent Spatio-Temporal}
\newacronym[shortplural=pdfs, longplural=probability density functions]{pdf}{pdf}{probability density function}
\newacronym[shortplural=CIs, longplural=Confidence Intervals]{ci}{CI}{Confidence Interval}

\glsenableentrycount 

\newcounter{daggerfootnote}

\usepackage[cachedir=minted-cache]{minted}
\usepackage{fancyvrb}
\usepackage[htt]{hyphenat}
\usepackage{url}
\usepackage{enumitem}

\newcommand{\x}{\ensuremath{\bm{x}}}
\newcommand{\y}{\ensuremath{\bm{y}}}
\newcommand{\z}{\ensuremath{\bm{z}}}
\newcommand{\q}{\ensuremath{\bm{q}}}

\newcommand{\norm}[1]{\ensuremath{\left\lVert #1 \right\rVert}}
\newcommand{\ang}[1]{\ensuremath{\left\langle #1 \right\rangle}}
\newcommand{\bigo}[1]{\ensuremath{\mathcal{O}\left( #1 \right)}}
\newcommand{\nt}{n_{\mathrm{tensors}}}

\newcommand{\footurl}[1]{\footnote{\url{#1}}}
\definecolor{codebackgroundcolor}{rgb}{0.95,0.95,0.95}
\setminted{
  breaklines=true,
  python3=true,
  bgcolor=codebackgroundcolor,
  autogobble,
  fontsize=\footnotesize
}

\setmintedinline{
  breaklines=true
}

\definecolor{tsnebg}{RGB}{250, 254, 254}

\newcommand{\systemname}{\gls{quiver}\xspace}
\newcommand{\bsbmname}{\gls{bsbmdvq}\xspace}
\newcommand{\dbpedianame}{\gls{dbpediams}\xspace}

\newcommand{\dbpedialeft}{NaturalPlace}
\newcommand{\dbpediaright}{Village}

\newcommand{\repourl}{https://anonymous.4open.science/r/Dense-KG-Vector-DB-03BB/README.md}
\newcommand{\faissurl}{https://anonymous.4open.science/r/faiss-9EF4/README.md}
\newcommand{\qleverurl}{https://anonymous.4open.science/r/QLever-9F62/README.md}
\newcommand{\rdftensorurl}{https://anonymous.4open.science/r/jena-datatensor-9B90/README.md}

\renewcommand{\repourl}{https://github.com/Dakantz/Dense-KG-Vector-DB}
\renewcommand{\faissurl}{https://github.com/Dakantz/faiss}
\renewcommand{\qleverurl}{https://github.com/Dakantz/QLever}
\renewcommand{\rdftensorurl}{https://github.com/Dakantz/jena-datatensor}
\newcommand{\sep}{, }
\newenvironment{keywords}{\emph{Keywords:}}{}

\DeclareRobustCommand\thanks[1]{\footnotemark
    \protected@xdef\@thanks{\@thanks
        \protect\footnotetext[\the\c@footnote]{#1}}
}
\def\tsc#1{\csdef{#1}{\textsc{\lowercase{#1}}\xspace}}
\tsc{WGM}
\tsc{QE}

\title{Efficient Dense Vector Search within Knowledge Graph Content Embeddings}
\author[1,*]{Benedikt Kantz}

\author[1]{Tobias Schreck}

\author[2]{Gianmaria Silvello}
\date{}

\affil[1]{Institute of Visual Computing, Graz University of Technology}
\affil[2]{Department of Information Engineering, University of Padua}
\affil[*]{Corresponding author: \texttt{benedikt.kantz@tugraz.at}}
\fancypagestyle{firstpage}{
  \fancyhf{}
}
\begin{document}

\let\WriteBookmarks\relax
\def\floatpagepagefraction{1}
\def\textpagefraction{.001}

\twocolumn
\maketitle

\begin{abstract}
  Knowledge graphs are a core component of today’s knowledge infrastructure, supporting reasoning and anchoring knowledge systems to verifiable facts. RDF stores and SPARQL engines fulfill this function, enabling a range of retrieval and inference tasks on structured knowledge. Coupling them with \cglspl{lm} extends RAG toward neurosymbolic reasoning, where structured queries gate or re-rank generative outputs. This line of reasoning requires that SPARQL evaluation natively support tensor operations on dense embeddings, enabling multimodal querying and learned similarity-based ranking to be expressed together with graph-structural constraints. This approach is feasible only if the engine can efficiently perform dense vector search.

  We present \systemname, an extension to QLever that adds native support for dense vector retrieval within RDF knowledge graphs. It implements three optimizations: engine-level registration of tensor functions, vocabulary-time parsing of JSON-encoded vectors, and a virtual \texttt{SERVICE} that exposes a vector index inside the query. We propose two new benchmarks: an extension of \gls{bsbm} with text embeddings and an extension of DBpedia with image embeddings. Against the baselines, vocabulary-time parsing alone yields median speedups of up to \bestSpeedupBsbmScanEmbedded on \gls{bsbm} and \bestSpeedupDbpediaScanEmbedded on DBpedia for single-type ranking; adding an approximate nearest-neighbor index yields speedups of \speedupBsbmBestScanIndexSystemnamebase on \gls{bsbm} and \bestSpeedupDbpediaScanIndex on DBpedia. The index further makes cross-modal vector joins on DBpedia feasible in seconds, whereas all non-indexed configurations time out.
\end{abstract}


\setlength{\columnsep}{3em}
\sloppy

\section{Introduction}
\label{sec:intro}

Search over knowledge-intensive data increasingly requires the joint evaluation of graph constraints and learned similarity. SPARQL evaluates graph patterns over RDF-modelled graphs (e.g., \glspl{kg}), including type constraints, property paths, and joins, whereas dense retrieval ranks entities or documents by proximity in an embedding space. Many multimodal retrieval and \gls{rag} workloads require both operations within a single query plan: vector similarity determines relevance, while graph patterns restrict the admissible results~\autocite{Abootorabi2025AskInAnyModality,Radford2021CLIP,Imran2026MultimodalVisionLanguageModels}. Vector retrieval alone can return candidates that fail the required graph constraints, while standard SPARQL provides no native operator for ranking bindings by embedding similarity.

Most of the current architectures handle the integration of graph pattern matching and vector similarity search through multi-stage pipelines: graph and vector components are implemented separately and composed sequentially, thus incurring additional data transfer overhead and hindering joint query optimisation. 

Among graph query engines, two approaches have been explored. The first is Neo4j which includes vector indexing as an integrated service, reducing the number of pipeline stages while retaining inter-system communication for vector operations. The second is RDFTensor, which provides the closest RDF-native operator interface, but its current design represents tensors textually and identifies faster lexical-to-value mapping as future work~\autocite{Marciniak2025DataTensorsRDF}. 

Database research studies this integration as hybrid query processing and, when structured predicates restrict the candidate vectors, filtered vector search~\autocite{Wei2020AnalyticDBV,Zhang2023VBASE,Chronis2025}. Recent systems extend this formulation to graph-query blocks~\autocite{Mohoney2023HighThroughputKG,Liu2025TigerVector} and to declarative model-backed operators~\autocite{Patel2025LOTUS,Liu2025Palimpzest}. However, these systems do not provide scalable native composition of dense vector operators with SPARQL graph patterns over RDF knowledge graphs.

We identify the cost of materializing a stored vector into a computable representation, repeated once per candidate at query time, as dominant in dense retrieval under SPARQL, rather than the similarity arithmetic that tuned linear-algebra kernels already handle at a fraction of that cost.

Addressing this limitation requires moving tensor decoding outside the query path, which is a structural change that cannot be retrofitted onto an existing query-time parsing model without modifying the storage layer.

We implement this restructuring, together with \gls{aknn} indexing, natively within QLever~\autocite{Bast2017QLever}, a state-of-the-art, open source, C++ triple store shown to outperform other off-the-shelf systems~\autocite{Bast2025Sparqloscope}. We target three specific optimisation levels:
\begin{enumerate}[label=$\alph*.$]
    \item \textbf{engine-level integration:} tensor operators are registered directly within the query execution pipeline, enabling joint optimisation with graph traversal and eliminating inter-system data transfer;
    \item \textbf{vocabulary-time parsing:} vectors are decoded from their serialized lexical form during index construction rather than at query execution, removing repeated parse overhead from the critical path of every query;

    \item \textbf{in-database \gls{aknn}:} a nearest-neighbour index built on the Faiss library~\autocite{Douze2025FaissLibrary} is exposed as a virtual \texttt{SERVICE} endpoint within the engine, replacing $\bigo{\nt}$ linear scans with sub-linear retrieval of $\bigo{\sqrt{\nt}}$ without leaving the graph query context.
\end{enumerate}
The proposed system, \textsc{\systemname}, is implemented as an extension to QLever~\autocite{Bast2017QLever}; it reuses the engine's vocabulary layer, permutation-index scans, and virtual \texttt{SERVICE} mechanism, and introduces a vector vocabulary, a tensor operator family, and a Faiss-backed \gls{aknn} service on top of these.
The introduced improvements make structurally new query types feasible for the first time, in particular multimodal joins that combine graph-structural constraints with cross-modal vector similarity in a single SPARQL query.

No existing benchmark exercises the queries this work targets: SPARQL conjunctions of type-membership and property-path constraints with vector-similarity ranking, and vector-similarity joins between typed entities. To isolate the contribution of each optimisation across dataset sizes, we extend the \gls{bsbm}~\autocite{Bizer2011BSBM} with sentence-transformer embeddings, scaling from $2.4\cdot 10^{3}$ to $3.5 \cdot 10^{7}$ triples generated automatically. To evaluate the same query family on real multimodal data, we extend DBpedia~\autocite{Lehmann2015DBpediaA} with CLIP~\autocite{Radford2021CLIP} embeddings of associated Wikimedia Commons images, producing a $4.2 \cdot 10^{8}$ triple benchmark for cross-modal similarity joins. These novel benchmark datasets are used to compare the existing RDFTensor implementation and Neo4j to our \systemname, demonstrating that our system is significantly faster across various settings, attesting to its efficiency and speed from small synthetic datasets up to real-world multimodal query settings.

In summary, we make the following contributions.
\begin{enumerate}[label=(\roman*)]
    
    \item We show through sub-timing profiling that query-time deserialization of serialized vector literals dominates SPARQL tensor processing and scales linearly in the number of tensors.
    \item We introduce \systemname, an open-source QLever extension enabling multimodal semantic querying while reducing vector search from linear to sub-linear running time within the SPARQL query plan.
    \item We present \bsbmname ($2.4\cdot10^{3}$ to $3.5\cdot10^{7}$ triples, text embeddings) and \dbpedianame ($4.2\cdot10^{8}$ triples, CLIP image embeddings), the first benchmarks for vector ranking and vector-similarity joins under SPARQL structural constraints.
    \item We evaluate \systemname against RDFTensor, Neo4j, and a two-stage pipeline, with significance testing: median speedups reach
    \bestSpeedupBsbmScanIndex on \bsbmname and \bestSpeedupDbpediaScanIndex on \dbpedianame, and cross-modal joins complete in seconds where all non-indexed configurations time out. Code, benchmarks, and harness are publicly released.
\end{enumerate}

The rest of the paper is organised as follows. Related systems are described in \Cref{sec:rel-works}. \Cref{sec:optim,sec:benchmarks} detail our optimisation strategy and evaluation framework, with \Cref{sec:impl} describing their implementation, with the code being available on \href{\repourl}{\repourl}. \Cref{sec:results} presents the experimental results, further discussed in \Cref{sec:disc}.

\section{Related Work}
\label{sec:rel-works}

Approaches to combining dense vector retrieval with structural graph querying differ along three axes: where the vector operator resides (external service or native in-engine), when encoded vectors are decoded (query time or ingestion time), and how a structural predicate composes with vector similarity (pre-, post-, or inline-filtering)~\autocite{Chronis2025}. We position \systemname against the literature, sweeping from the most modular architectures to fully in-engine integrations.

\paragraph{Standalone \glspl{vdbms} and federated pipelines}

Standalone \glspl{vdbms} maintain modularity but incur inter-process overhead when coupled with a structural query engine. Specialized systems such as Milvus, Pinecone, and Qdrant expose \gls{aknn} interfaces over large vector collections and, in some cases, embed inputs directly within the engine~\autocite{Taipalus2024VectorDBMS,Pan2024VectorDatabase}. Their expressiveness, however, is limited relative to general \gls{dbms}~\autocite{Taipalus2024VectorDBMS}, and integrating them into retrieval systems or graph query engines reverts to a multi-stage architecture: either large intermediate results are transferred between execution contexts (foregoing joint optimisation), or search accuracy is sacrificed because the full vector range cannot be scanned, posing an effectiveness/efficiency trade-off~\autocite{Asadi2013EffectivenessEfficiencyMultiStageRetrieval,Sheng2025MQRLD}. Multimodal \gls{kg} construction pipelines that compose vision and language models with structural retrieval follow the same federated pattern and inherit the same coupling cost~\autocite{Lee2024MultimodalReasoningKG,Kannan2020MultimodalKGCodePapers,Bai2026SepRelationAwareKGLearning}. The advent of cross-encoders for retrieval across modalities~\autocite{Abootorabi2025AskInAnyModality,Zhao2024DenseTRSurvey,Tian2026TrajectorySimilarityJoin,Gu2024HPSTopKTrajectory} makes this overhead dominant in end-to-end query latency, motivating native integration.

\paragraph{Native vector integration in relational and graph DBMS}
Native integration of vector operators inside a \gls{dbms} eliminates inter-process communication overhead but presents another bottleneck: repeated query-time decoding of vector literals. In the relational domain, AnalyticDB-V fused vector and structured-query execution within a single analytical engine, treating \gls{aknn} indices as physical operators under accuracy-aware cost-based optimisation~\autocite{Wei2020AnalyticDBV}. VBASE generalised this design by identifying \emph{relaxed monotonicity} as a common property that lets vector and relational operators share an execution interface beyond the top-$k$ contract~\autocite{Zhang2023VBASE}. SingleStore-V extended the pattern to a distributed in-memory \gls{dbms}~\autocite{Chen2024SingleStoreV}, while \texttt{pgvector} \footurl{https://github.com/pgvector/pgvector} provides a comparable extension to PostgreSQL.

In the graph domain, \textcite{Marciniak2025DataTensorsRDF} implemented a range of operators as a Jena extension~\autocite{Carroll2004JenaPaper} together with a new \acrshort{rdf} data type for tensors. Their solution, RDFTensor, can perform search operations over a \gls{kg},

but the Java-based implementation suffers from two inefficiencies: the underlying engine fails to scale to large-scale \glspl{kg}~\autocite{Bast2025Sparqloscope}, with the authors acknowledging that the implementation is too slow at parsing tensors to operate within feasible time bounds on large data. Neo4j integrates a vector index as an internal service but restricts its use to top-level lookups,\footurl{https://neo4j.com/docs/cypher-manual/current/indexes/semantic-indexes/vector-indexes} and TigerVector adds \gls{aknn} acceleration to a property-graph engine over the GSQL query language~\autocite{Liu2025TigerVector}.
Across this lineage, native integration emerges as the common denominator for joint optimisation; closing the \emph{parse-time bottleneck} that remains in the graph branch is the gap our work addresses. Integrated dense search is, notably, distinct from graph embeddings~\autocite{Wang2023SurveyHeterogeneousGraphEmbedding}, which are extracted from the structure of the \gls{kg} rather than stored within it. \acrshort{rdf}-side proposals along that direction~\autocite{Ristoski2019RDF2VecRDFGraphEmbeddings,Cochez2017GlobalRDFVectorSpaceEmbeddings,Kulmanov2018Vec2SPARQL} encode triples for downstream tasks, whereas we optimise operations on integrated vectors as a first-class data type. \systemname does not preclude folding those representations \emph{back into the graph} for joint querying.

\paragraph{\gls{fvs} and hybrid query optimisation}
When a structural predicate restricts the candidate set, the \gls{aknn} problem becomes \gls{fvs}, formalised by~\textcite{Chronis2025} into a pre-, post-, and inline-filtering taxonomy with associated stable-recall objectives. ACORN augments an \gls{hnsw} index with predicate-subgraph traversal to support arbitrary predicates without compromising recall~\autocite{Patel2024ACORN}; Filtered-DiskANN constructs a disk-resident graph index whose edges are aware of the label set associated with each vector, supporting equality predicates at scale~\autocite{Gollapudi2023FilteredDiskANN}; SeRF addresses range-filter variants by maintaining a segment graph over ordered attributes~\autocite{Zuo2024SeRF}. The join queries we evaluate are an instance of \gls{fvs} in which the filter is expressed as a SPARQL graph pattern rather than a tabular predicate, and our virtual \texttt{SERVICE} design corresponds to an inline-filtering execution method in this taxonomy.

\paragraph{Modality extensions to SPARQL}
Modality-specific extensions to SPARQL provide the methodological precedent for in-engine integration. The spatial domain has produced both multi-stage operators~\autocite{Shi2016TopKRelevantSemanticPlaceRetrieval} and integrated approaches~\autocite{Battle2011GeoSPARQL,Jovanovik2021GeoSPARQLComplianceBenchmark,Bast2025SpatialJoins}; text search beyond filter techniques has followed the same dichotomy, with two-stage~\autocite{PerezAguera2010UsingBM25FSemanticSearch,Dosso2020SearchTextRetrieveGraphs,Blanco2011EntitySearchRDFData} and integrated variants~\autocite{Bast2017QLever}. The integrated approaches reuse the same engine-level pattern, that is, a typed literal, a custom operator, and a virtual \texttt{SERVICE} call when indexing applies; however, none handle variations in user-supplied vector encodings, and none support dense vectors.

Within this landscape, \systemname occupies the inline-filtering quadrant of \gls{fvs} at the storage layer of an open-source SPARQL engine: tensor literals are decoded once at ingestion time, vector operators are first-class within the query execution pipeline, and an \gls{aknn} index is exposed as a virtual \texttt{SERVICE} that participates in joint planning with graph traversal. These optimisations reduce both time spent per tensor operation, and reduce the scaling over vector search to $\bigo{\sqrt{n}}$.

This implementation closes the parse-time bottleneck that limits the only prior in-engine \acrshort{rdf} attempt~\autocite{Marciniak2025DataTensorsRDF}, while reusing the architectural pattern that the same engine already applies to spatial and text modalities~\autocite{Bast2025SpatialJoins,Bast2017QLever}.

\section{\acrshort{quiver} System Design}
\label{sec:optim}

\tikzexternaldisable
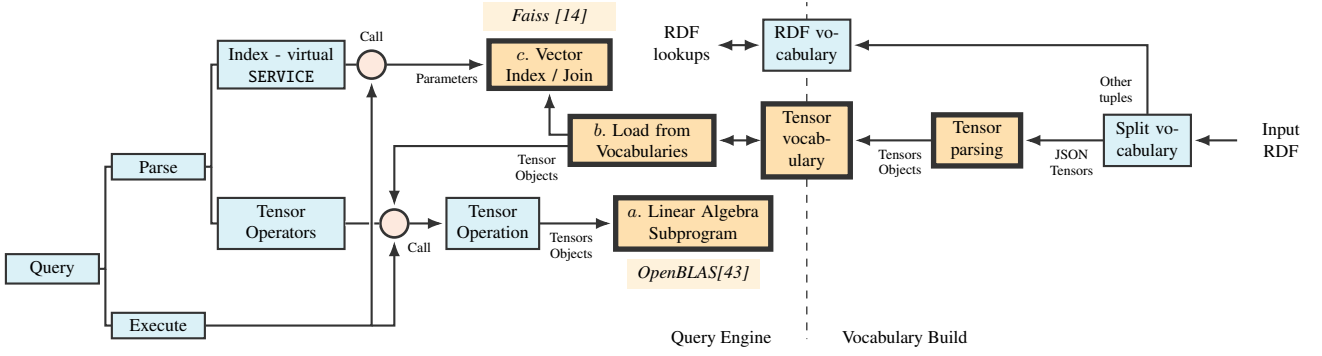
\begin{figure*}
    \centering
    \newcommand{\nodesep}{1.8cm}
\newcommand{\nodesepl}{0.2cm}
\newcommand{\halfnodesep}{0.8cm}
\newcommand{\lblsep}{0.09em}
\newcommand{\citesep}{0.4em}
\newcommand{\lblsepx}{-0.2em}
\newcommand{\nodeabove}{1.8cm}
\newcommand{\dualoff}{0.1cm}

\begin{tikzpicture}[
    scale=0.75, transform shape,
    glow/.style={preaction={draw, line join=round, opacity=0.9,line width=2pt,#1, shorten <=1pt, shorten >=1pt}},
    glow/.default=yellow,
    archelem/.style={rectangle, text width=4em, fill=SkyBlue!20, draw=black!80, font={\small}, text centered, line width=0.8pt, anchor=center},
    archpath/.style={draw=black!80, line width=0.8pt, glow=white},
    archpathb/.style={->, arrows={-{Latex[length=2mm]}}, draw=black!80, line width=0.8pt, glow=white, shorten >=1pt},
    archpathr/.style={<-, archpathb},
    archpathboth/.style={<->, archpathb, arrows={{Latex[length=2mm]}-{Latex[length=2mm]}}},
    archlbl/.style={fill=none,  font={\small\it}, text centered},
    optimlbl/.style={archlbl, fill=Dandelion!20, opacity=0.8, text opacity=1},
    archelem_highlight/.style={archelem,fill=Dandelion!40, line width=2pt},
    archexec/.style={circle, text width=0.2cm, fill=Melon!20, draw=black!80, align=center, font={\small}, line width=1pt},
    coord/.style={circle, text width=0cm,},
    desclblb/.style={archlbl, font={\scriptsize}},
    inout_lbl/.style={archlbl, font={\small}},
    ]

    \pgfdeclarelayer{background}
    \pgfdeclarelayer{foreground}
    \pgfsetlayers{background,main,foreground}

    \node (q) [archelem] {Query};
    \node (parse) [above right=\nodeabove and \nodesepl of q.east, archelem, anchor=west] {Parse};
    \node (exec) [below right=1cm and \nodesepl of q.east, archelem, anchor=west] {Execute};
    \node (service) [above right=\nodeabove and \nodesepl of parse.east, archelem, text width=2cm, anchor=west] {Index - virtual \texttt{SERVICE}};
    \node (operators) [below right=1cm and \nodesepl of parse.east, archelem, text width=2cm, anchor=west] {Tensor Operators};

    \node (faiss) [right=(\nodesep+\nodesep) of service.east, archelem_highlight, text width=1.9cm]{$c.$ Vector Index / Join};
    \node (faiss_lbl) [above=\citesep of faiss, optimlbl, text width=6em]{Faiss \cite{Douze2025FaissLibrary}};
    \node (faiss_lbl_in) [below left=\lblsep and \lblsepx of faiss.west, desclblb, text width=1.3cm]{Parameters};

    \node (vector_op) [right=(\nodesep + \halfnodesep) of operators, archelem, text width=1.4cm]{Tensor Operation};
    \node (vector_eng) [right=1.3cm of vector_op.east, archelem_highlight, text width=2.5cm, anchor=west]{$a.$ Linear Algebra Subprogram};
    \node (vector_eng_lbl) [below=\citesep of vector_eng, optimlbl, text width=6em]{OpenBLAS\cite{Wang2013OpenBLAS}};
    \node (vector_eng_lbl_in) [below right=\lblsep and \lblsepx of vector_op.east, desclblb, text width=1cm]{Tensors Objects};
    
    \node (load_vocab) [above right=1cm and \nodesep of vector_op, archelem_highlight, text width=2.3cm]{$b.$ Load from Vocabularies};

    \node (vocab) [right=\halfnodesep of load_vocab, archelem_highlight, text width=1.3cm, anchor=west]{Tensor vocabulary};
    \node (parse_tensor) [right=2.2cm of vocab, archelem_highlight, text width=1.3cm]{Tensor parsing};
    \node (split_vocab) [right=2.2cm of parse_tensor, archelem, text width=1.3cm]{Split vocabulary};
    \node (rdf_vocab) [above=1cm of vocab, archelem, text width=1.3cm]{RDF vocabulary};
    \node (rdf_in) [right=\halfnodesep of split_vocab, inout_lbl, text width=1.3cm, anchor=west]{Input \acrshort{rdf}};

    \node (vocab_out) [left=\halfnodesep of rdf_vocab.west, inout_lbl, text width=1cm, anchor=east]{\acrshort{rdf} lookups};

    \node (split_vocab_out_vec) [below left=\lblsep and \lblsepx of split_vocab.west, desclblb, text width=1cm]{JSON Tensors};
    \node (split_vocab_out_rdf) [above left=\lblsep and \lblsepx of split_vocab.north, desclblb, text width=1cm]{Other tuples};
    \node (parse_out_vec) [below left=\lblsep and \lblsepx of parse_tensor.west, desclblb, text width=1cm]{Tensors Objects};

    \node (load_out_lbl) [below left=\lblsep+\dualoff and \lblsepx of load_vocab.west, desclblb, text width=1cm]{Tensor Objects};

    \node (op_service) [right=0.2cm of service, archexec]{};
    \node (op_op)   [right=0.6cm of operators, archexec]{};

    \node (op_lbl) [below right=\lblsep and \lblsepx of op_op, desclblb]{Call};
    \node (service_lbl) [above=\lblsep of op_service, desclblb]{Call};

    \node (q_l_) at ($(q.east)!0.5!(exec.west)$ ){};
    \node (q_l) at (q_l_ |- q.east){};
    \node (p_l_) at ($(parse.east)!0.5!(operators.west)$ ){};
    \node (p_l) at (p_l_ |- parse.east){};
    \path[archpath] (q.east) -- (q_l.center) |- (parse);
    \path[archpath] (q_l.center) |- (exec);
    \path[archpath] (parse.east) -- (p_l.center) |- (service);
    \path[archpath] (p_l.center) |- (operators);

    \path[archpath] (service.east) -- (op_service);
    \path[archpathb] (op_service) -- (faiss);
    \path[archpathb] (operators.east) -- (op_op) -- (vector_op);

    \path[archpathboth] (vocab) -- (load_vocab);
    \path[archpathb] ($(load_vocab.west)+(0,+\dualoff)$) -|  (faiss);
    \path[archpathr]  ($(load_vocab.west)-(0,+\dualoff)$ -| faiss) -| (op_op);
    \path[archpathb] (vector_op) -- (vector_eng);
    \path[archpathb] (exec.east) -| (op_service);
    \path[archpathb] (exec.east -| op_service) -| (op_op);

    \path[archpathb] (rdf_in) -- (split_vocab);
    \path[archpathb] (split_vocab) -- (parse_tensor);
    \path[archpathb] (parse_tensor) -- (vocab);
    
    \path[archpathb] (split_vocab) |-  (rdf_vocab);
    \path[archpathboth] (vocab_out) -- (rdf_vocab);
    \coordinate (vocab_south)  at (vocab.south |- exec.south);
    \begin{pgfonlayer}{background}
        \path[draw,dashed] (vocab.north) -- (vocab.north |- faiss_lbl.north);
        \path[draw,dashed] (vocab.south) -- (vocab_south);
    \end{pgfonlayer}
    \node (q_lbl) [inout_lbl, left=0.5cm of vocab_south, anchor=east] {Query Engine};
    \node (v_lbl) [inout_lbl, right=0.5cm of vocab_south, anchor=west] {Vocabulary Build};

\end{tikzpicture}
    \caption{Overview of \systemname's three optimisations across QLever's query parsing and execution flow. We highlight our core integrations and optimisations in orange. They are $a.$ integrating OpenBLAS~\autocite{Wang2013OpenBLAS} directly, $b.$ leveraging our pre-built tensor vocabulary, and, finally, $c.$ integrating Faiss~\autocite{Douze2025FaissLibrary} directly in the database through a virtual \texttt{SERVICE}~\autocite{Bast2025SpatialJoins}. }
    \label{fig:engine_arch}
\end{figure*}
\tikzexternalenable

\systemname integrates three optimisations into the QLever SPARQL engine~\autocite{Bast2017QLever,Bast2025Sparqloscope} at successive levels of abstraction: operator-level functions, vocabulary-time decoding, and engine-level indexing. The implementation supports arbitrary-dimensional tensors; the practical use cases focus on order-one tensors (vectors).

QLever has demonstrated both high efficiency and high performance, even when handling large, real-world datasets such as DBpedia and Wikidata. It is implemented in modern C++ and makes extensive use of contemporary language features to design the engine from the ground up. We build our features and optimisations directly into the engine itself. 
QLever follows a principled \gls{dbms} approach for \acrshort{rdf} storage and query processing~\autocite{Cure2015RDFDatabaseSystems}. \Cref{fig:engine_arch} reports the main components of \systemname.
Starting from the right side of \Cref{fig:engine_arch}, we observe that the system first constructs a \emph{vocabulary} from the literals appearing in the input tuples. At this stage, the engine, by default, does not distinguish between different data types. As we explain below, this characteristic opens up potential avenues for optimisation.

After this, the system constructs the permutation-based index: by default, two permutations (PSO and POS) over the integer-encoded triples, optionally all six variants can be built~\autocite{Bast2017QLever}.

Once the vocabulary and the permutation indices have been built, the engine is ready to serve queries; the corresponding execution path is shown on the left side of \Cref{fig:engine_arch}. At startup, \systemname registers two interfaces with the QLever planner. The tensor operators listed in \Cref{tab:impl_operators} are exposed as SPARQL extension functions identified by 
\acrshort{uri}s in the \texttt{dtf:} namespace, so they can appear inside any \texttt{BIND}, \texttt{FILTER}, or projection clause of an incoming query. The \gls{aknn} index is exposed as a virtual \texttt{SERVICE} endpoint, reusing the same mechanism that QLever already employs for spatial joins~\autocite{Bast2025SpatialJoins}. When a query arrives, the parser produces a query plan in which calls to a \texttt{dtf:} function and to the virtual \texttt{SERVICE} are ordinary operator nodes. During execution, the engine retrieves the corresponding inputs from the vocabulary, decodes them when necessary, and dispatches the call to the extension's code path.

Execution of the query plan proceeds through QLever's standard execution pipeline: scans over the permutation indices, joins, and projections. When execution reaches a node corresponding to a \texttt{dtf:} function or to the virtual \texttt{SERVICE}, the engine resolves the argument variables, retrieves the bound tensor literals from the vocabulary, converts each into the internal tensor representation, and passes the decoded tensors to the function registered for that IRI. In the default path this conversion parses the JSON-encoded payload at every invocation, which is the dominant operator overhead the following subsections address.

\paragraph{$a.$ Linear Algebra Subprogram}

The operator-level integration (component $a.$ in \Cref{fig:engine_arch}) is \systemname's
baseline configuration.
Linear-algebra primitives are dispatched to OpenBLAS~\autocite{Wang2013OpenBLAS}, which provides tuned implementations on commodity CPU architectures, statically linked into QLever. A second implementation path, written against the C++ standard library, is compiled in as a fallback for platforms on which OpenBLAS is not available. The operators exposed at this layer are listed in \Cref{tab:impl_operators}: they cover the arithmetic and similarity primitives required for dense-vector retrieval and are a deliberate subset of the operator set defined by RDFTensor~\autocite{Marciniak2025DataTensorsRDF}, since the remaining RDFTensor operations are not used by the retrieval workloads we target~\autocite{Zhao2024DenseTRSurvey}.
\begin{table*}[tb]\footnotesize
    \centering
    \caption{Implemented operators. $\x\in\mathbb{R}^d$ and $\y\in\mathbb{R}^d$ represent input tensors of the type \texttt{dtf:DataTensor} with $d$ dimensions, $\z\in\mathbb{R}^d$ an output tensor. These operators establish the basic mathematical functions required to perform search over a collection of vectors.}
    \label{tab:impl_operators}
    \setlength\tabcolsep{0.5em}
\begin{tabular}{lllll}
    \toprule
    Operator                       & Inputs     & Output & Function                                  & Note                            \\
    \midrule
    \texttt{dtf:tensor}            & string     & $\z$   & parsing                                   & Parses a string to a tensor.    \\
    \texttt{dtf:norm2}             & $\x$       & float  & $\norm{\x}_2 $                            & Compute the $\ell_2$ norm.      \\
    \texttt{dtf:add}               & $\x$, $\y$ & $\z$   & $\x +\y $                                 & Add the two tensors.            \\
    \texttt{dtf:subtract}          & $\x$, $\y$ & $\z$   & $\x -\y $                                 & Subtract the two tensors.       \\
    \texttt{dtf:dotProduct}        & $\x$, $\y$ & float  & $\ang{x,y}    $                           & Compute the dot product.        \\
    \texttt{dtf:cosineSimilarity}  & $\x$, $\y$ & float  & $ \frac{\ang{x,y}}{\norm{x}_2\norm{y}_2}$ & Compute the cosine similarity.  \\
    \texttt{dtf:euclideanDistance} & $\x$, $\y$ & float  & $\norm{\x-\y}_2 $                         & Compute the Euclidian distance. \\
    \bottomrule
\end{tabular}
\end{table*}

These operations, by design, require already-parsed tensors for their processing. The tensors within the tuples are, however, stored in a human-readable format to enable the interpretation of the data. 
We decode data from the existing \acrshort{json} format using the RapidJSON\footurl{https://rapidjson.org} library for fast parsing. This is applied to all incoming tensor literals, whether they come from the query, the stored vocabulary ($b.$), or other computations.

\paragraph{$b.$ Vector Vocabularies}

Because the tensors stored in vocabularies are known in advance of any user-submitted queries, we define a dedicated database vocabulary for these vectors. This shifts the parsing of tensor data from query execution time to the vocabulary construction phase, which in turn decreases execution-time loading overhead.
This optimised loading mechanism is already in use for other multimodal data in QLever, where essential geometric properties are extracted from spatial structures to accelerate queries during runtime~\autocite{Bast2025SpatialJoins}. Similarly, our method processes the vectors during triple indexing by identifying all tokens within the triples that are of tensor type, parsing them with the previously mentioned \acrshort{json} library, and then storing them in a binary format using a memory-mapped lookup table. A secondary lookup table is required to deal with the arbitrary lengths and dimensions of vectors. The remaining tokens in the triples are directly passed through to the default vocabularies.

This vocabulary can subsequently be used both by the operations to load the tensors required for computation and by the indexed system ($c.$).

\paragraph{$c.$ Vector Index / Join}

The optimisations introduced above reduce the constant factor per tensor visited but leave the asymptotic complexity of similarity search unchanged: comparing a query vector against every stored tensor is still linear in the number of tensors ($\nt$) in the query's scope. Such a scan returns the $k$-nearest neighbours of a query vector $\q \in \mathbb{R}^d$ among the indexed vectors $\x_i \in \mathbb{R}^d$ under a similarity measure $f_{\mathrm{sim}}(\q, \x_i)$~\autocite{Zhao2024DenseTRSurvey}.
Sub-linear retrieval is achievable through \gls{aknn} indices, which trade exact recall for reduced lookup cost by exploiting the spatial structure of the vector space, either via vector quantisation or via graph-based traversal~\autocite{Bernhardsson2018Annoy,Douze2025FaissLibrary,Hajebi2011KNNGraph,Li2019ANNHighDimensionalData,Zhao2024DenseTRSurvey}.

In practice, \gls{aknn} indices achieve sub-linear retrieval cost, $\bigo{\sqrt{\nt}}$ empirically~\autocite{Douze2025FaissLibrary}. We use Faiss~\autocite{Douze2025FaissLibrary}, a widely used library for efficient dense indexed retrieval~\autocite{Wang2022DenseExternalExpansion, Li2027DualTowerMultimodalEntityLinking, Goyal2025QFASKE, Yang2026TextToCypherPipeline}, in our indexed implementation. Two index families dominate: \gls{hnsw} maintains a hierarchical navigable graph traversed greedily from coarse to fine layers, while \gls{ivf} partitions the vector space into Voronoi cells via coarse $k$-means clustering and routes each query to its closest cells. The choice between them depends on dataset size, memory budget, and the recall--latency operating point required by the application;
\textcite{Douze2025FaissLibrary} recommend \gls{ivf} at scale and emphasise that parameters of either index must be tuned to the target vector distribution and workload.
\begin{table*}
    \centering
    \caption{Overview of the indexing parameters. 
    }
    \label{tab:faiss-parameter}
    \setlength\tabcolsep{0.5em}

\begin{tabular}{llp{0.4\linewidth}ll}
    \toprule
    Index type                     & Parameter              & Description                                                               & Key                 & Default      \\
    \midrule
    \multirow[c]{4}{*}{all}        & $k$                    & Number of nearest neighbors to retrieve                                   & \texttt{numNN}      & 100          \\
    \cline{2-5}
                                   & Index                  & Search algorithm to use. (\texttt{naive}, \texttt{hnsw}, or \texttt{ivf}) & \texttt{algorithm}  & \texttt{ivf} \\
    \cline{2-5}
                                   & Distance               & Search algorithm to use. (\texttt{dot} or \texttt{euclidian})             & \texttt{distance}   & \texttt{dot} \\
    \cline{1-5}
    \multirow[c]{3}{*}{\gls{hnsw}} & $M$                    & Number of edges for each in the index graph                               & \texttt{nNeighbors} & $\sqrt{\nt}$ \\
    \cline{2-5}
                                   & $ef_{\mathrm{search}}$ & Expansion factor for search, i.e. depth to probe                          & \texttt{searchK}    & 16           \\
    \cline{1-5}
    \multirow[c]{2}{*}{\gls{ivf}}  & $K_{\mathrm{IVF}}$     & Number of lists used to build index                                       & \texttt{kIVF}       & $\sqrt{\nt}$ \\
    \cline{2-5}
                                   & $n_{\mathrm{probe}}$   & Lists to visit during search                                              & \texttt{searchK}    & 1            \\
    \bottomrule
\end{tabular}
\end{table*}
\systemname exposes the \gls{aknn} search parameters to the user; \Cref{tab:faiss-parameter} lists those that govern the recall--latency trade-off, with the unified key \texttt{searchK} mapping to $n_{\mathrm{probe}}$ for \gls{ivf} and to $ef_{\mathrm{search}}$ for \gls{hnsw}.
The defaults favour latency, and in \Cref{sec:results} we measure the trade-off across the range of \texttt{searchK} values. The \gls{aknn} index is exposed as a virtual SPARQL \texttt{SERVICE} endpoint: QLever intercepts \texttt{SERVICE} blocks whose IRI matches the extension's namespace and routes them to internal code rather than to a remote endpoint, reusing the mechanism that already powers QLever's spatial-join service~\autocite{Bast2025SpatialJoins}. A call binds the \gls{aknn} operands to two argument variables, sets index parameters (\texttt{numNN}, \texttt{searchK}, distance metric) via blank-node properties, and projects neighbour identifiers (and, when requested, the corresponding distances) into the outer query plan. \Cref{lis:tensor-index} shows an example that uses default values for the remaining parameters.

\begin{listing}[tb]
    \centering

    \caption{A join query asking \enquote{For each \texttt{dbo:\dbpedialeft}, find the closest \texttt{dbo:\dbpediaright} with regard to their thumbnail embedding} where we perform a call to the virtual \texttt{SERVICE} endpoint.}
    \begin{Verbatim}[fontsize=\small]
SELECT * WHERE {
    ?l a dbo:NaturalPlace .
    ?l dbo:thumbnail_embedding ?left_emb .
    # Call to external index 
    SERVICE tensorIndex: { 
        # Specify the left embedding to join on
        _:config tensorIndex:left ?left_emb ; 
        # Specify the right embedding to join on
        tensorIndex:right ?right_emb ; 
        # We just want one closest neighbour
        tensorIndex:numNN 1 . 
        { # specify the right-hand subquery
            ?r a dbo:Village ;
                dbo:thumbnail_embedding ?right_emb .
        }
    }
}
    \end{Verbatim}
    \label{lis:tensor-index}
\end{listing}

The intercepted call is dispatched to \systemname's \gls{aknn} module, which builds a Faiss index on demand from the parameters specified in the \texttt{SERVICE} block~\autocite{Douze2025FaissLibrary}. On-demand construction is required because the vector set is determined by the SPARQL subquery and is therefore query-dependent; a single index built once over the full collection would, on a query whose surrounding graph pattern restricts the candidate set, produce a neighbourhood structure tuned to the full distribution. Such a structure degrades recall on the filtered subset, a phenomenon documented in the literature~\autocite{Chronis2025,Patel2024ACORN,Gollapudi2023FilteredDiskANN}. Once built, each index is held together with its input vectors in a size-bounded in-memory cache and evicted when the cache budget is exceeded; the cache covers the index structure but not the SPARQL subquery that produced the vector set, so repeated invocations re-execute the subquery unless QLever's query cache already holds the result.

\section{Experimental Setup}
\label{sec:benchmarks}

Assessing multimodal query engines requires large \glspl{kg} augmented with dense entity embeddings. Earlier work has examined how to connect Wikimedia resources with DBpedia, but has primarily relied on strict semantic properties rather than similarity in a vector space~\autocite{Ferrada2017IMGpedia,Wang2020Richpedia}. Spatial data~\autocite{Bast2021EfficientRDFConverterOSM} and video~\autocite{Vizcarra2021HumanBehaviorIndexingVideo} have been linked similarly. Other approaches, as outlined in \Cref{sec:rel-works}, use multiple parallel pipelines~\autocite{Lee2024MultimodalReasoningKG,Kannan2020MultimodalKGCodePapers}, falling into the two-stage systems category.

To our knowledge, no prior benchmark exists for tensor operations within SPARQL query engines. The proposed benchmark comprises a synthetic dataset for scalability stress tests, a real-world dataset for multimodal retrieval, and a framework that measures query latency and resource consumption.

We remark that benchmark does not evaluate the retrieval performance of the vector representations or \gls{aknn} indices, but focuses on efficiency and feasibility of the proposed queries.

\subsection{Datasets}

\paragraph{\acrfull{bsbmdvq}}

\bsbmname extends \gls{bsbm}~\autocite{Bizer2011BSBM}, the standard synthetic SPARQL scalability benchmark, by attaching a dense-vector encoding to every \acrshort{rdf} comment and label. Each text field is encoded with a \texttt{sentence-transformer} model~\autocite{Reimers2019SBERT} and serialised in the \acrshort{json} format defined by RDFTensor~\autocite{Marciniak2025DataTensorsRDF}; the embedding is attached via the original predicate suffixed with \verb|_embedding|, leaving the \gls{bsbm} schema otherwise untouched. \gls{bsbm} labels are random word concatenations, and the resulting embeddings therefore carry no semantic structure; this is acceptable because \bsbmname stresses engine throughput rather than retrieval quality.

We inherit \gls{bsbm}'s scale parameter $m$, producing the six size points in \Cref{tab:bsbm_counts} that span $5.8 \cdot 10^2$ to $3.1 \cdot 10^5$ tensor literals over $2.4 \cdot 10^3$ to $3.5 \cdot 10^7$ total triples.

\begin{table}[tb]
    \centering
    \caption{Tuple counts within the \bsbmname~\autocite{Bizer2011BSBM} sets after encoding. $m$ represents the argument used for generating the tuples (\enquote{requested tuples}), $n$ the total number of tuples including dense vectors, $n_{\mathrm{tensors}}$ the number of tuples containing a dense vector.}
    \setlength\tabcolsep{0.5em}
\begin{tabular}{lrrr}
    \toprule
    Power & $m$      & $n$                & $\nt$ \\
    \midrule
    0     & $10^{0}$ & $2.4 \cdot 10^{3}$ & $5.8 \cdot 10^{2}$     \\
    1     & $10^{1}$ & $5.6 \cdot 10^{3}$ & $6.1 \cdot 10^{2}$     \\
    2     & $10^{2}$ & $4.2 \cdot 10^{4}$ & $2.2 \cdot 10^{3}$     \\
    3     & $10^{3}$ & $3.8 \cdot 10^{5}$ & $1.2 \cdot 10^{4}$     \\
    4     & $10^{4}$ & $3.6 \cdot 10^{6}$ & $4.2 \cdot 10^{4}$     \\
    5     & $10^{5}$ & $3.5 \cdot 10^{7}$ & $3.1 \cdot 10^{5}$     \\
    \bottomrule
\end{tabular}

    \label{tab:bsbm_counts}
\end{table}

\paragraph{DBpedia for Multimodal Search}
\dbpedianame extends as DBpedia~\autocite{Lehmann2015DBpediaA} with image embeddings, mirroring the construction of \bsbmname but replacing the \texttt{sentence-transformer} with a multimodal encoder. Each DBpedia subject is resolved to its Wikipedia article through the \path{foaf:isPrimaryTopicOf} predicate; the associated thumbnail is retrieved from the \cgls{wit} corpus~\autocite{Srinivasan2021WIT}, encoded with CLIP~\autocite{Radford2021CLIP}, and serialised under the original predicate suffixed with \verb|_embedding| in the same \acrshort{json} format as in \bsbmname. The resulting graph contains $2.36 \cdot 10^6$ tuples with dense vectors among the $4.22 \cdot 10^8$ triples of the English DBpedia release. \dbpedianame supports cross-modal joins that combine a SPARQL graph pattern with image-similarity ranking.

\subsection{Queries}

\begin{figure*}
    \centering
    \usetikzlibrary {shadows}
\begin{tikzpicture}[
    table_elements/.style={
            text width=\nodewidth, align=center,
            minimum height=\nodeheight, anchor=center
        },
    table_defs/.style={
            table_elements, minimum height=\diffheight, font={\bf}, anchor=south
        },
    table_defs_drive/.style={
            table_defs,text width=\drivewidth, minimum height=\nodeheight, anchor=east
        },
    qnode/.style={rectangle, text width=0.3cm, fill=qnode!20, draw=black!80, align=center, font={\tiny}},
    qqueryvec/.style={qnode, fill=Goldenrod!20, text width=0.3cm},
    qnode_con/.style={thick,  arrows={-{Latex[length=2mm]}}, shorten >=0.2mm},
    q_con/.style={thick,  arrows={-{Latex[length=1.5mm]}}, shorten >=0.2mm},
    q_con_lbl/.style={fill=white, font={\tiny}, anchor=center, pos=0.35},
    dist_measure/.style={fill=SpringGreen!10, draw=black!80, font={\tiny}, anchor=center},
    dist_measure_lbl/.style={font={\tiny}, anchor=center},
    table_desc/.style={text width=\nodewidth, align=center,font={\small}},
    q_desc/.style={text width=\descwidth, align=center,font={\tiny\it}, draw=black!30, fill=black!5, line width=2pt, anchor=north, align=left},
    emb_in/.style={font={\small},  align=center},
    emb_link/.style={line width=1pt,  arrows={-{Latex[length=1.5mm]}}, draw=black!50},
    emb_linklbl/.style={fill=white, font={\tiny}, anchor=south, pos=0.45, inner sep=2pt },
    transparency group
    ]

    \newcommand{\nodeheight}{5cm}
    \newcommand{\nodeheightscan}{4cm}
    
    \newcommand{\nodewidth}{0.38\textwidth}
    \newcommand{\descwidth}{0.32\textwidth}
    \newcommand{\diffheight}{1.2em}
    \newcommand{\diffoff}{3em}
    \newcommand{\qoff}{2.5em}
    \newcommand{\qspane}{1em}
    \newcommand{\qspanh}{2.7em}
    \newcommand{\embsep}{0.9cm}
    \newcommand{\drivewidth}{2.5em}
    \definecolor{bsbm}{rgb}{0.8,0.2,0.4}

    \definecolor{dbpedia}{rgb}{0.2,0.5,0.2}

    \definecolor{scan}{rgb}{0.3,0.8,0.3}
    \definecolor{join}{rgb}{0.8,0.3,0.3}

    \pgfdeclarelayer{background}
    \pgfdeclarelayer{foreground}
    \pgfsetlayers{background,main,foreground}
    \definecolor{qnode}{rgb}{0.3,0.3,0.9}
    \definecolor{qnodelink}{rgb}{0.1,0.2,0.5}
    \matrix[
        matrix of nodes, nodes in empty cells,
        nodes={table_elements,
                draw=black!20},
    ]{
        \node (scan_bsbm) [minimum height=\nodeheightscan] {}; & \node (scan_dbpedia) [minimum height=\nodeheightscan] {}; \\
        \node (join_bsbm) {}; & \node (join_dbpedia) {}; \\
    };

    \node (scan_bsbm_q_desc) [below=0.3em of scan_bsbm.north, q_desc] {$k$ subjects that have any \texttt{bsbmv:productFeature} and are most similar to \texttt{?qv}.  };
    \node (scan_bsbm_q_p) [below left=0em and \qspane of scan_bsbm.center, qnode] {?p};
    \node (scan_bsbm_q_v) [below=\qoff of scan_bsbm_q_p, qnode] {?v};
    \node (scan_bsbm_q_f) [above=\qoff of scan_bsbm_q_p, qnode] {?f};

    \draw [->,q_con] (scan_bsbm_q_p) -- node [q_con_lbl] {\texttt{bsbmv:productFeature}} (scan_bsbm_q_f);
    \draw [->,q_con] (scan_bsbm_q_p) -- node [q_con_lbl] {\texttt{rdf:comment\_embedding}} (scan_bsbm_q_v);
    
    \node (scan_bsbm_e) [left=\embsep of scan_bsbm_q_v, emb_in] {\faFileTextO};
    \draw [->,emb_link] (scan_bsbm_e) -- node [emb_linklbl] {embed} (scan_bsbm_q_v);

    \node (scan_bsbm_q_qv) [right=(\qspane + \qspane) of scan_bsbm_q_v, qqueryvec] {?qv};

    \begin{pgfonlayer}{background}
        \node (scan_bsbm_dist_q) [fit=(scan_bsbm_q_qv) (scan_bsbm_q_v), dist_measure]    {};
        \node (scan_bsbm_dist_q_lbl) [below=0.5em of scan_bsbm_dist_q, dist_measure_lbl] {\texttt{BIND dtf:dotProduct(?v, ?qv) AS ?dist}};
    \end{pgfonlayer}

    \node (scan_dbpedia_q_desc) [below=0.3em of scan_dbpedia.north, q_desc] {$k$ \texttt{dbo:\dbpedialeft}s that have a vector \texttt{dbo:thumbnail\_embedding} and are most similar to \texttt{?qv}.  };
    \node (scan_dbpedia_q_p) [below left=0em and \qspane of scan_dbpedia.center, qnode] {?l};
    \node (scan_dbpedia_q_v) [below=\qoff of scan_dbpedia_q_p, qnode] {?v};
    \node (scan_dbpedia_q_f) [above=\qoff of scan_dbpedia_q_p, qnode, text width=5em] {dbo:\dbpedialeft};

    \draw [->,q_con] (scan_dbpedia_q_p) -- node [q_con_lbl] {\texttt{a}} (scan_dbpedia_q_f);
    \draw [->,q_con] (scan_dbpedia_q_p) -- node [q_con_lbl] {\texttt{dbo:thumbnail\_embedding}} (scan_dbpedia_q_v);

    \node (scan_dbpedia_q_qv) [right=(\qspane + \qspane) of scan_dbpedia_q_v, qqueryvec] {?qv};

    \node (scan_dbpedia_e) [left=\embsep of scan_dbpedia_q_v, emb_in] {\faFilePhotoO};
    \draw [->,emb_link] (scan_dbpedia_e) -- node [emb_linklbl] {embed} (scan_dbpedia_q_v);

    \begin{pgfonlayer}{background}
        \node (scan_dbpedia_dist_q) [fit=(scan_dbpedia_q_qv) (scan_dbpedia_q_v), dist_measure]    {};
        \node (scan_dbpedia_dist_q_lbl) [below=0.5em of scan_dbpedia_dist_q, dist_measure_lbl] {\texttt{BIND dtf:dotProduct(?v, ?qv) AS ?dist}};
    \end{pgfonlayer}

    \node (join_bsbm_q_desc) [below=0.3em of join_bsbm.north, q_desc] {$k$ subjects that have any \texttt{bsbmv:productFeature} and their \emph{one} most similar \texttt{rdf:comment\_embedding} to one subject that has any \texttt{bsbmv:productFeature}.};

    \node (join_bsbm_q_p1) [below left=0em and \qspanh of join_bsbm.center, qnode] {?p1};
    \node (join_bsbm_q_v1) [below=\qoff of join_bsbm_q_p1, qnode] {?v1};
    \node (join_bsbm_q_f1) [above=\qoff of join_bsbm_q_p1, qnode] {?f1};

    \node (join_bsbm_q_p2) [right=(\qspanh+\qspanh) of join_bsbm_q_p1.east, qnode] {?p2};
    \node (join_bsbm_q_v2) [below=\qoff of join_bsbm_q_p2, qnode] {?v2};
    \node (join_bsbm_q_f2) [above=\qoff of join_bsbm_q_p2, qnode] {?f2};

    \draw [->,q_con] (join_bsbm_q_p1) -- node [q_con_lbl] {\texttt{bsbmv:productFeature}} (join_bsbm_q_f1);
    \draw [->,q_con] (join_bsbm_q_p1) -- node [q_con_lbl] {\texttt{rdf:comment\_embedding}} (join_bsbm_q_v1);
    \draw [->,q_con] (join_bsbm_q_p2) -- node [q_con_lbl] {\texttt{bsbmv:productFeature}} (join_bsbm_q_f2);
    \draw [->,q_con] (join_bsbm_q_p2) -- node [q_con_lbl] {\texttt{rdf:comment\_embedding}} (join_bsbm_q_v2);

    \node (join_bsbm_e1) [left=\embsep of join_bsbm_q_v1, emb_in] {\faFileTextO};
    \draw [->,emb_link] (join_bsbm_e1) -- node [emb_linklbl] {embed} (join_bsbm_q_v1);
    \node (join_bsbm_e2) [right=\embsep of join_bsbm_q_v2, emb_in] {\faFileTextO};
    \draw [->,emb_link] (join_bsbm_e2) -- node [emb_linklbl] {embed} (join_bsbm_q_v2);

    \begin{pgfonlayer}{background}
        \node (join_bsbm_dist_q) [fit=(join_bsbm_q_v1) (join_bsbm_q_v2), dist_measure]    {};
        \node (join_bsbm_dist_q_lbl) [below=0.5em of join_bsbm_dist_q, dist_measure_lbl] {\texttt{BIND dtf:dotProduct(?v1, ?v2) AS ?dist}};
        \node (join_bsbm_dist_q_lbl_2) [below=0.2em of join_bsbm_dist_q_lbl, dist_measure_lbl] {ORDER BY ?dist LIMIT 1};
    \end{pgfonlayer}

    \node (join_bsbm_q_desc) [below=0.3em of join_dbpedia.north, q_desc] {$k$ \texttt{dbo:\dbpedialeft}s and their \emph{one} most similar \texttt{dbo:thumbnail\_embedding} to one \texttt{dbo:\dbpediaright}. };

    \node (join_dbpedia_q_p1) [below left=0em and \qspanh of join_dbpedia.center, qnode] {?l};
    \node (join_dbpedia_q_v1) [below=\qoff of join_dbpedia_q_p1, qnode] {?v1};
    \node (join_dbpedia_q_f1) [above=\qoff of join_dbpedia_q_p1, qnode, text width=5em] {dbo:\dbpedialeft};

    \node (join_dbpedia_q_p2) [right=(\qspanh+\qspanh) of join_dbpedia_q_p1.east, qnode] {?r};
    \node (join_dbpedia_q_v2) [below=\qoff of join_dbpedia_q_p2, qnode] {?v2};
    \node (join_dbpedia_q_f2) [above=\qoff of join_dbpedia_q_p2, qnode, text width=5em] {dbo:\dbpediaright};

    \draw [->,q_con] (join_dbpedia_q_p1) -- node [q_con_lbl] {\texttt{a}} (join_dbpedia_q_f1);
    \draw [->,q_con] (join_dbpedia_q_p1) -- node [q_con_lbl] {\texttt{dbo:thumbnail\_embedding}} (join_dbpedia_q_v1);
    \draw [->,q_con] (join_dbpedia_q_p2) -- node [q_con_lbl] {\texttt{a}} (join_dbpedia_q_f2);
    \draw [->,q_con] (join_dbpedia_q_p2) -- node [q_con_lbl] {\texttt{dbo:thumbnail\_embedding}} (join_dbpedia_q_v2);

    \node (join_dbpedia_e1) [left=\embsep of join_dbpedia_q_v1, emb_in] {\faFilePhotoO};
    \draw [->,emb_link] (join_dbpedia_e1) -- node [emb_linklbl] {embed} (join_dbpedia_q_v1);
    \node (join_dbpedia_e2) [right=\embsep of join_dbpedia_q_v2, emb_in] {\faFilePhotoO};
    \draw [->,emb_link] (join_dbpedia_e2) -- node [emb_linklbl] {embed} (join_dbpedia_q_v2);

    \begin{pgfonlayer}{background}
        \node (join_dbpedia_dist_q) [fit=(join_dbpedia_q_v1) (join_dbpedia_q_v2), dist_measure]    {};
        \node (join_dbpedia_dist_q_lbl) [below=0.5em of join_dbpedia_dist_q, dist_measure_lbl] {\texttt{BIND dtf:dotProduct(?v1, ?v2) AS ?dist}};
        \node (join_dbpedia_dist_q_lbl_2) [below=0.2em of join_dbpedia_dist_q_lbl, dist_measure_lbl] {ORDER BY ?dist LIMIT 1};
    \end{pgfonlayer}

    \node (bsbm) at (scan_bsbm.north) [table_defs, fill=bsbm!50,draw=bsbm!50] {\bsbmname};
    \node (dbpedia) at (scan_dbpedia.north)[table_defs, fill=dbpedia!50, draw=dbpedia!50] {\dbpedianame};

    \node (scan) at (scan_bsbm.west) [table_defs_drive, fill=scan!50, draw=scan!50,minimum height=\nodeheightscan] {Scan};
    \node (join) at (join_bsbm.west) [table_defs_drive, fill=join!50, draw=join!50] {Join};

\end{tikzpicture} 
    \caption{Overview of our four query possibilities across the \bsbmname / \dbpedianame datasets and scan / join queries. Each query is limited to $k=10$ distinct \emph{output} tuples and ordered by the distance \texttt{?dist}. The vector \texttt{?qv} for the scan queries is provided at query time and supplied by the same models we use to generate the embeddings, while the internal embeddings are generated before we initialize the databases. Textual descriptions for each query explain the semantic details.}
    \label{fig:query-possibilities}
\end{figure*}

Both \bsbmname and \dbpedianame are evaluated with the same two query templates, so the evaluation harness is reused without modification. Each template computes vector similarity through the \texttt{dtf:dotProduct} operator over the stored dense vectors and ranks results by the resulting dot product. The two templates are illustrated in \Cref{fig:query-possibilities} and referred to throughout as \emph{scan} and \emph{join} queries.

\begin{description}
    \item[Scan.] A $k$-nearest-neighbour ranking over the entities of a single common type relative to a query vector $\bm{q}$ supplied at runtime. A linear scan over every type-matching entity yields $\bigo{\nt}$ time; an \gls{aknn} index reduces this to $\bigo{\sqrt{\nt}}$, in line with the empirical Faiss bound~\autocite{Douze2025FaissLibrary}.
    \item[Join.] A vector-similarity join between entities of two common types: for each entity of the left type, the most similar entity of the right type is returned. The lookup is repeated for every left-type entity, giving $\bigo{\nt^2}$ time under a linear scan; with the \gls{aknn} index the cost drops to $\bigo{\nt\sqrt{\nt}}$.
\end{description}

The scan and join queries are applied to the respective domains, while maintaining a comparable structure and complexity. In addition, we retrieve only distinct tuples and sort them by distance. For better performance, the indexed queries replace the \texttt{dtf:dotProduct} operator. These indexed queries employ the \gls{ivf} method for \gls{aknn}, using the default search parameters with $k=10$. Any further scan and join operation would, while increasing complexity, add another expected factor in computation cost, hence producing no further insights for our contributions, yielding no further insights into our contributed system.

The \dbpedianame set enables us to place additional constraints on the query, mimicking how a real-world semantic scan and join query can be constructed. As already shown throughout \Cref{lis:tensor-index,fig:query-possibilities}, we use \texttt{dbo:\dbpedialeft} and \texttt{dbo:\dbpediaright} as our class constraints, reducing the overall search space for the dense vector query. This constraint effectively scopes the scan to $46.3 \cdot 10^3$ tuples, and the join over the cross product $\text{\texttt{dbo:\dbpedialeft}} \times \text{\texttt{dbo:\dbpediaright}}$ to a total of $1.44 \cdot 10^9$ comparisons in the unoptimised case.

\subsection{Baselines}

Beyond RDFTensor, the evaluation includes two additional baselines that capture the federated and property-graph architectures discussed in \Cref{sec:rel-works}. Finally, the \gls{aknn} search is performed with the Python bindings of Faiss~\autocite{Douze2025FaissLibrary}. This pipeline emulates how a dedicated \gls{vdbms}~\autocite{Abootorabi2025AskInAnyModality} would interact with a SPARQL engine that lacks native vector support. It is run only on \bsbmname, since the inter-process transfer cost already dominates at moderate scale (see \Cref{sec:results}).

\emph{Neo4j.} The property-graph comparator is Neo4j running its native vector index. Two caveats apply. First, the \acrshort{rdf}-to-property-graph mapping is itself a design choice that influences the resulting graph structure and downstream query timings~\autocite{Angles2020MappingRDFDatabasesToPropertyGraphDatabases,Khayatbashi2022ConvertingPropertyGraphsToRDF}, so a single-point comparison does not isolate the index from the encoding. Second, Neo4j's vector index supports only an \gls{aknn}-first execution path, in which the similarity search runs before any structural filter. \footurl{https://neo4j.com/docs/cypher-manual/current/clauses/search/\#limitations} The join queries evaluated in this paper require the opposite order, namely a structural filter first followed by a similarity search restricted to the filtered subset, so they cannot be executed inside Neo4j's index machinery. Both datasets are nevertheless loaded into Neo4j through the Neosemantics plugin\footurl{https://neo4j.com/labs/neosemantics/}, a plugin used to ingest \acrshort{rdf} triples into a \gls{pg} system. This allows \systemname and Neo4j to be compared on \bsbmname for overall scalability, with limited comparability in exact timings.

\subsection{Evaluation Harness}

The evaluation runs every $\langle\text{system}, \text{query}, \text{dataset}\rangle$ combination across the four \systemname configurations (\systemname-Base, \systemname-TV, \systemname-Base+Index, \systemname-TV+Index), the RDFTensor baseline on Fuseki~\autocite{Marciniak2025DataTensorsRDF}, and the two baselines introduced above: the two-stage pipeline and Neo4j (only on \bsbmname). Of these baselines, RDFTensor is the closest in terms of architectural compatibility, as the queries are even portable to our \systemname, thus serving as the benchmark reference. Any other choice would have led to less comparability due to differences in graph representations (Neo4j with property graphs and our triple store), database principles (relational \gls{vdbms} exposed by a graph abstraction), or pipelines with multiple stages (large parameter space, different execution frameworks).

Reliable wall-clock measurements on database engines are, furthermore, non-trivial because cache hierarchies, operating-system scheduling, and hardware specifics introduce variance that single-shot timings cannot capture~\autocite{Currim2016DBMSMetrology}. Each combination is therefore repeated up to 128 times, with the run aborted earlier if two 30-second timeouts occur; the median over the completed runs is reported~\autocite{Crolotte2009BenchmarkMetricSelection}, together with CPU and memory consumption recorded over the same window. Cache effects are mitigated across repetitions by varying the query vector (whenever the template admits one) and the SPARQL query text, thereby defeating string-keyed query-result caches in the evaluated engines. The harness additionally measures the sub-timings to investigate our query optimisiations by logging the relevant timepoints, allowing us to trace the individual query through the systems.

\paragraph{Statistical Tests}

Statistical claims use a significance threshold of $\alpha = 0.01$ with the Bonferroni correction applied per dataset and per query template over the family of pairwise comparisons~\autocite{Lehmann2022TestingStatisticalHypotheses}, shown in the respective results tables. We perform a one-sided t-test with unequal variance for all timing comparisons, yielding the $p$-values for the query times compared to the reference baseline.

\section{Implementation \& Evaluation Environment}
\label{sec:impl}

\systemname consists of two parts: the engine extension itself and the benchmark suite used to evaluate it. The engine extension is implemented directly within the existing C++ implementation of QLever; the RapidJSON and Faiss libraries are linked directly into the engine, requiring no further dynamic loading. The implementation contains unit- and integration tests to ensure correct functionality of the operators, new vocabularies, and indices, and allow further extensions while keeping regressions to a minimum.

Both dataset extensions are constructed in Python with the \texttt{sentence-transformers} library~\autocite{Reimers2019SBERT}. \bsbmname encodes \acrshort{rdf} comments and labels with the \texttt{all-MiniLM-L6-v2} model into a 384-dimensional vector space. \dbpedianame encodes Wikimedia Commons thumbnails with the multimodal \texttt{clip-ViT-B-32} model~\autocite{Radford2021CLIP}, exposed through the same library. The smaller CLIP variant is selected for its low per-image encoding cost; more recent multimodal encoders~\autocite{Abootorabi2025AskInAnyModality} would yield higher retrieval quality at proportionally higher embedding-construction cost -- for which we do not benchmark in this work.
The evaluation harness is also implemented in Python, where we manage the databases' processes, monitoring their status and resource requirements while the queries are executed by them. To access the query engines, we used HTTP requests sent by the \texttt{rdflib} library~\autocite{RDFLib2023}, with any processing only performed after calculating the query timings. These are taken as close to the result retrieval as possible, with no engine-dependent logic between. The sub-timings are acquired from log points within the engines, recompiled for just the sub-timings tests.
All queries were executed and performance metrics were measured in series on a machine with one \enquote{AMD Ryzen 9 9900X 12-Core} processor with 92 gigabytes of memory. The whole suite requires approximately 650 gigabytes of storage.

\paragraph{Reproducibility}
\label{sec:suppl}

The \systemname source code is available on Anonymous GitHub, \footurl{\qleverurl} as is the benchmark suite, including the code to recreate the datasets \footurl{\repourl}.
Source code for minor adaptations to Faiss and the RDFTensor library, enabling our benchmark suite, is also available on Anonymous GitHub. \footurl{\rdftensorurl}\textsuperscript{,}\footurl{\faissurl} \dbpedianame is available via Zenodo,\footurl{https://zenodo.org/records/19664692} but can be reproduced using the above repositories.

\begin{figure*}[tb]
    \centering
    \begin{subfigure}{0.48\textwidth}
        \centering
        \input{figures/generated/overview/bsbm_timings_Scan_legend_full}
        \caption{Scan query.}
        \label{fig:bsbm_timings_easy}
    \end{subfigure}
    \hfill
    \begin{subfigure}{0.48\textwidth}
        \centering
        \input{figures/generated/overview/bsbm_timings_Join_legend_full}
        \caption{Join query.}
        \label{fig:bsbm_timings_hard}
    \end{subfigure}

    \caption{Median timings for the scan (\subref{fig:bsbm_wall_easy}) and join (\subref{fig:bsbm_timings_hard}) query on \bsbmname as the total tuple count $n$ grows. The baseline (RDFTensor~\autocite{Marciniak2025DataTensorsRDF}) and Neo4j are compared against \systemname with and without the tensor vocabulary (TV) and with and without the AkNN index. Error bars indicate the 5th and 95th percentiles; both axes are logarithmic. Dotted lines indicate polynomial extrapolations.}
    \label{fig:bsbm_timings}
\end{figure*}

\begin{figure*}[tb]
    \centering

    \begin{tikzpicture}[
scale=0.5,
every axis/.style={
                    legend pos=south west, 
                    legend style={
                        font=\small},
                    legend columns=1, 
                }
]

\definecolor{brown1765533}{RGB}{176,55,33}
\definecolor{dodgerblue0143213}{RGB}{0,143,213}
\definecolor{lightgrey203}{RGB}{203,203,203}
\definecolor{lightgrey204}{RGB}{204,204,204}
\definecolor{teal0100149}{RGB}{0,100,149}
\definecolor{tomato2527948}{RGB}{252,79,48}
\definecolor{whitesmoke240}{RGB}{240,240,240}

\begin{axis}[
axis line style={whitesmoke240},
height=0.54\linewidth,
legend cell align={left},
legend style={
  fill opacity=0.8,
  draw opacity=1,
  text opacity=1,
  at={(0.03,0.97)},
  anchor=north west,
  draw=lightgrey204
},
tick align=outside,
tick pos=left,
title={Index Query Performance},
width=0.9\linewidth,
x grid style={lightgrey203},
xlabel={Recall@16},
xmajorgrids,
xmin=0.08125, xmax=1.04375,
xtick style={color=black},
xtick={0,0.2,0.4,0.6,0.8,1,1.2},
xticklabels={
  \(\displaystyle {0.0}\),
  \(\displaystyle {0.2}\),
  \(\displaystyle {0.4}\),
  \(\displaystyle {0.6}\),
  \(\displaystyle {0.8}\),
  \(\displaystyle {1.0}\),
  \(\displaystyle {1.2}\)
},
y grid style={lightgrey203},
ylabel={Time \(\displaystyle t\) (s)},
ymajorgrids,
ymin=0.038, ymax=0.054,
ytick style={color=black},
ytick={0.038,0.04,0.042,0.044,0.046,0.048,0.05,0.052,0.054},
yticklabels={
  \(\displaystyle {0.038}\),
  \(\displaystyle {0.040}\),
  \(\displaystyle {0.042}\),
  \(\displaystyle {0.044}\),
  \(\displaystyle {0.046}\),
  \(\displaystyle {0.048}\),
  \(\displaystyle {0.050}\),
  \(\displaystyle {0.052}\),
  \(\displaystyle {0.054}\)
}
]
\addplot [thick, dodgerblue0143213, mark=*, mark size=3, mark options={solid}]
table {
0.125 0.0414108037948608
0.25 0.0417760610580444
0.375 0.0432797670364379
0.5 0.0445705652236938
0.5625 0.0428617000579833
0.5625 0.042286992073059
0.6875 0.0441516637802124
0.6875 0.0457018613815307
0.6875 0.0439903736114502
};
\addlegendentry{HNSW}
\path [draw=dodgerblue0143213, thick]
(axis cs:0.125,0.0394040942192077)
--(axis cs:0.125,0.0437459945678711);

\path [draw=dodgerblue0143213, thick]
(axis cs:0.25,0.0385449528694153)
--(axis cs:0.25,0.0447954535484313);

\path [draw=dodgerblue0143213, thick]
(axis cs:0.375,0.0413862466812133)
--(axis cs:0.375,0.0482789874076843);

\path [draw=dodgerblue0143213, thick]
(axis cs:0.5,0.0408100485801696)
--(axis cs:0.5,0.0488120317459106);

\path [draw=dodgerblue0143213, thick]
(axis cs:0.5625,0.0397681593894958)
--(axis cs:0.5625,0.0473018288612365);

\path [draw=dodgerblue0143213, thick]
(axis cs:0.5625,0.0397596359252929)
--(axis cs:0.5625,0.0449784994125366);

\path [draw=dodgerblue0143213, thick]
(axis cs:0.6875,0.0400303602218627)
--(axis cs:0.6875,0.0475382804870605);

\path [draw=dodgerblue0143213, thick]
(axis cs:0.6875,0.0415107607841491)
--(axis cs:0.6875,0.0483986139297485);

\path [draw=dodgerblue0143213, thick]
(axis cs:0.6875,0.040198266506195)
--(axis cs:0.6875,0.0474105477333068);

\addplot [thick, dodgerblue0143213, opacity=1, mark=-, mark size=5, mark options={solid}, only marks, forget plot]
table {
0.125 0.0394040942192077
0.25 0.0385449528694153
0.375 0.0413862466812133
0.5 0.0408100485801696
0.5625 0.0397681593894958
0.5625 0.0397596359252929
0.6875 0.0400303602218627
0.6875 0.0415107607841491
0.6875 0.040198266506195
};
\addplot [thick, dodgerblue0143213, opacity=1, mark=-, mark size=5, mark options={solid}, only marks, forget plot]
table {
0.125 0.0437459945678711
0.25 0.0447954535484313
0.375 0.0482789874076843
0.5 0.0488120317459106
0.5625 0.0473018288612365
0.5625 0.0449784994125366
0.6875 0.0475382804870605
0.6875 0.0483986139297485
0.6875 0.0474105477333068
};
\addplot [thick, tomato2527948, mark=square*, mark size=3, mark options={solid}]
table {
0.1875 0.043030858039856
0.3125 0.0425242185592651
0.5625 0.0439543724060058
0.6875 0.0448588132858276
0.8125 0.045701265335083
0.875 0.0469125509262085
0.9375 0.0485279560089111
1 0.0484163761138915
1 0.0476688146591186
};
\addlegendentry{IVF}
\path [draw=tomato2527948, thick]
(axis cs:0.1875,0.0401906371116638)
--(axis cs:0.1875,0.0467532873153686);

\path [draw=tomato2527948, thick]
(axis cs:0.3125,0.0399160981178283)
--(axis cs:0.3125,0.0450799465179443);

\path [draw=tomato2527948, thick]
(axis cs:0.5625,0.0414355397224426)
--(axis cs:0.5625,0.0484173297882079);

\path [draw=tomato2527948, thick]
(axis cs:0.6875,0.0418643951416015)
--(axis cs:0.6875,0.0479992032051086);

\path [draw=tomato2527948, thick]
(axis cs:0.8125,0.0425491333007812)
--(axis cs:0.8125,0.0529800653457641);

\path [draw=tomato2527948, thick]
(axis cs:0.875,0.0446503162384033)
--(axis cs:0.875,0.0508193373680114);

\path [draw=tomato2527948, thick]
(axis cs:0.9375,0.0455803275108337)
--(axis cs:0.9375,0.05301034450531);

\path [draw=tomato2527948, thick]
(axis cs:1,0.0457578301429748)
--(axis cs:1,0.0518217086791992);

\path [draw=tomato2527948, thick]
(axis cs:1,0.0444381833076476)
--(axis cs:1,0.0510515570640563);

\addplot [thick, dodgerblue0143213, opacity=1, mark=-, mark size=5, mark options={solid,draw=tomato2527948}, only marks, forget plot]
table {
0.1875 0.0401906371116638
0.3125 0.0399160981178283
0.5625 0.0414355397224426
0.6875 0.0418643951416015
0.8125 0.0425491333007812
0.875 0.0446503162384033
0.9375 0.0455803275108337
1 0.0457578301429748
1 0.0444381833076476
};
\addplot [thick, dodgerblue0143213, opacity=1, mark=-, mark size=5, mark options={solid,draw=tomato2527948}, only marks, forget plot]
table {
0.1875 0.0467532873153686
0.3125 0.0450799465179443
0.5625 0.0484173297882079
0.6875 0.0479992032051086
0.8125 0.0529800653457641
0.875 0.0508193373680114
0.9375 0.05301034450531
1 0.0518217086791992
1 0.0510515570640563
};
\draw (axis cs:0.125,0.0414108037948608) ++(-5pt,5pt) node[
  anchor=south east,
  text=teal0100149,
  rotate=0.0
]{1};
\draw (axis cs:0.25,0.0417760610580444) ++(-5pt,5pt) node[
  anchor=south east,
  text=teal0100149,
  rotate=0.0
]{4};
\draw (axis cs:0.375,0.0432797670364379) ++(-5pt,5pt) node[
  anchor=south east,
  text=teal0100149,
  rotate=0.0
]{8};
\draw (axis cs:0.5625,0.0428617000579833) ++(-5pt,5pt) node[
  anchor=south east,
  text=teal0100149,
  rotate=0.0
]{16};
\draw (axis cs:0.6875,0.0439903736114502) ++(-5pt,5pt) node[
  anchor=south east,
  text=teal0100149,
  rotate=0.0
]{32};
\draw (axis cs:0.1875,0.043030858039856) ++(-5pt,5pt) node[
  anchor=south east,
  text=brown1765533,
  rotate=0.0
]{1};
\draw (axis cs:0.3125,0.0425242185592651) ++(-5pt,5pt) node[
  anchor=south east,
  text=brown1765533,
  rotate=0.0
]{2};
\draw (axis cs:0.5625,0.0439543724060058) ++(-5pt,5pt) node[
  anchor=south east,
  text=brown1765533,
  rotate=0.0
]{4};
\draw (axis cs:0.8125,0.045701265335083) ++(-5pt,5pt) node[
  anchor=south east,
  text=brown1765533,
  rotate=0.0
]{8};
\draw (axis cs:1,0.0476688146591186) ++(-5pt,5pt) node[
  anchor=south east,
  text=brown1765533,
  rotate=0.0
]{16};
\end{axis}

\end{tikzpicture}
    \caption{Median timings for the scan query on the \bsbmname of $m=10^4$ compared to the Recall\@@16. We compare \gls{hnsw} with \gls{ivf} over a sweep of \texttt{searchK} (indicated by the text besides the points) on a warm cache. The error bars indicate the quartiles.}
    \label{fig:bsbm_timings_index}
\end{figure*}
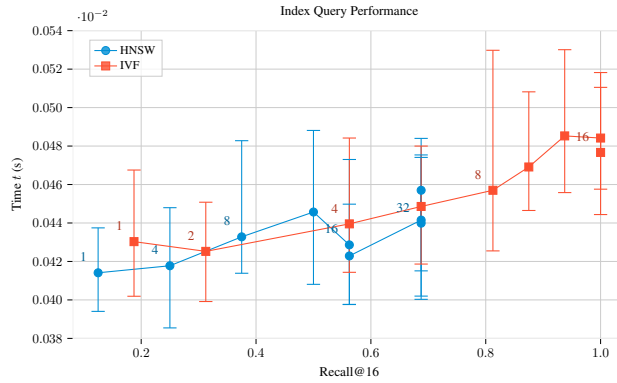

\begin{figure*}[tb]
    \centering
    \begin{subfigure}{0.48\textwidth}
        \centering
        \begin{tikzpicture}[
scale=0.5,
every axis/.style={
                        legend pos=south west, 
                        legend style={
                    font=\small},
                        legend columns=5, 
                        extra x ticks={200,300,400,500,600,700,800,900,2000,3000,4000,5000,6000,7000,8000,9000,20000,30000,40000,50000,60000,70000,80000,90000,200000,300000,400000,500000,600000,700000,800000,900000,2000000,3000000,4000000,5000000,6000000,7000000,8000000,9000000,20000000,30000000,40000000,50000000,60000000,70000000,80000000,90000000,200000000,300000000,400000000,500000000,600000000,700000000,800000000,900000000},
                        extra x tick labels={ },
                        extra y tick labels={ },
                        extra tick style={
                            tick style={draw=none},
                            major grid style={dashed,red},
                            grid=major,
                        },
                    }
]

\definecolor{darkgoldenrod17612723}{RGB}{176,127,23}
\definecolor{darkolivegreen7610055}{RGB}{76,100,55}
\definecolor{dodgerblue0143213}{RGB}{0,143,213}
\definecolor{goldenrod22917456}{RGB}{229,174,56}
\definecolor{lightgrey203}{RGB}{203,203,203}
\definecolor{lightgrey204}{RGB}{204,204,204}
\definecolor{olivedrab10914479}{RGB}{109,144,79}
\definecolor{tomato2527948}{RGB}{252,79,48}
\definecolor{whitesmoke240}{RGB}{240,240,240}

\begin{axis}[
axis line style={whitesmoke240},
height=1.2\linewidth,
legend cell align={left},
legend columns=2,
legend style={
  fill opacity=0.8,
  draw opacity=1,
  text opacity=1,
  at={(0.03,0.97)},
  anchor=north west,
  draw=lightgrey204
},
log basis x={10},
log basis y={10},
tick align=outside,
tick pos=left,
title={ BSBM-DVQ CPU Time \(\displaystyle t_c\) (s) (log scale) for Scan queries},
width=2\linewidth,
x grid style={lightgrey203},
xlabel={Dataset Size (log scale)},
xmajorgrids,
xmin=1497.32325667246, xmax=56808268.1898843,
xminorgrids,
xmode=log,
xtick style={color=black},
xtick={100,1000,10000,100000,1000000,10000000,100000000,1000000000},
xticklabels={
  \(\displaystyle {10^{2}}\),
  \(\displaystyle {10^{3}}\),
  \(\displaystyle {10^{4}}\),
  \(\displaystyle {10^{5}}\),
  \(\displaystyle {10^{6}}\),
  \(\displaystyle {10^{7}}\),
  \(\displaystyle {10^{8}}\),
  \(\displaystyle {10^{9}}\)
},
y grid style={lightgrey203},
ylabel={CPU Time \(\displaystyle t_c\) (s) (log scale)},
ymajorgrids,
ymin=0.001, ymax=55.0214998245918,
yminorgrids,
ymode=log,
ytick style={color=black},
ytick={0.0001,0.001,0.01,0.1,1,10,100,1000},
yticklabels={
  \(\displaystyle {10^{-4}}\),
  \(\displaystyle {10^{-3}}\),
  \(\displaystyle {10^{-2}}\),
  \(\displaystyle {10^{-1}}\),
  \(\displaystyle {10^{0}}\),
  \(\displaystyle {10^{1}}\),
  \(\displaystyle {10^{2}}\),
  \(\displaystyle {10^{3}}\)
}
]
\addplot [ultra thick, darkolivegreen7610055, dash pattern=on 4pt off 5pt, mark=triangle*, mark size=3, mark options={solid}]
table {
2418 0
5595 0
42417 0.01
383751 0.02
3577077 0.0300000000000002
35177974 0.16
};
\addlegendentry{\systemname-TV + index}
\path [draw=darkolivegreen7610055]
(axis cs:2418,0)
--(axis cs:2418,0.01);

\path [draw=darkolivegreen7610055]
(axis cs:5595,0)
--(axis cs:5595,0.01);

\path [draw=darkolivegreen7610055]
(axis cs:42417,0)
--(axis cs:42417,0.02);

\path [draw=darkolivegreen7610055]
(axis cs:383751,0.0099999999999997)
--(axis cs:383751,0.02);

\path [draw=darkolivegreen7610055]
(axis cs:3577077,0.0299999999999993)
--(axis cs:3577077,0.04);

\path [draw=darkolivegreen7610055]
(axis cs:35177974,0.15)
--(axis cs:35177974,0.1865);

\addplot [darkolivegreen7610055, mark=-, mark size=5, mark options={solid}, only marks, forget plot]
table {
2418 0
5595 0
42417 0
383751 0.0099999999999997
3577077 0.0299999999999993
35177974 0.15
};
\addplot [darkolivegreen7610055, mark=-, mark size=5, mark options={solid}, only marks, forget plot]
table {
2418 0.01
5595 0.01
42417 0.02
383751 0.02
3577077 0.04
35177974 0.1865
};
\addplot [ultra thick, darkgoldenrod17612723, dash pattern=on 4pt off 5pt, mark=diamond*, mark size=3, mark options={solid}]
table {
2418 0
5595 0
42417 0.01
383751 0.02
3577077 0.0300000000000002
35177974 0.16
};
\addlegendentry{\systemname-Base + index}
\path [draw=darkgoldenrod17612723]
(axis cs:2418,0)
--(axis cs:2418,0.01);

\path [draw=darkgoldenrod17612723]
(axis cs:5595,0)
--(axis cs:5595,0.01);

\path [draw=darkgoldenrod17612723]
(axis cs:42417,0)
--(axis cs:42417,0.02);

\path [draw=darkgoldenrod17612723]
(axis cs:383751,0.0099999999999997)
--(axis cs:383751,0.0299999999999999);

\path [draw=darkgoldenrod17612723]
(axis cs:3577077,0.0299999999999993)
--(axis cs:3577077,0.04);

\path [draw=darkgoldenrod17612723]
(axis cs:35177974,0.149999999999999)
--(axis cs:35177974,0.180000000000001);

\addplot [darkgoldenrod17612723, mark=-, mark size=5, mark options={solid}, only marks, forget plot]
table {
2418 0
5595 0
42417 0
383751 0.0099999999999997
3577077 0.0299999999999993
35177974 0.149999999999999
};
\addplot [darkgoldenrod17612723, mark=-, mark size=5, mark options={solid}, only marks, forget plot]
table {
2418 0.01
5595 0.01
42417 0.02
383751 0.0299999999999999
3577077 0.04
35177974 0.180000000000001
};
\addplot [ultra thick, olivedrab10914479, mark=triangle*, mark size=3, mark options={solid}]
table {
2418 0
5595 0
42417 0.01
383751 0.02
3577077 0.109999999999999
35177974 0.850000000000001
};
\addlegendentry{\systemname-TV}
\path [draw=olivedrab10914479]
(axis cs:2418,0)
--(axis cs:2418,0.01);

\path [draw=olivedrab10914479]
(axis cs:5595,0)
--(axis cs:5595,0.01);

\path [draw=olivedrab10914479]
(axis cs:42417,0)
--(axis cs:42417,0.02);

\path [draw=olivedrab10914479]
(axis cs:383751,0.01)
--(axis cs:383751,0.0300000000000002);

\path [draw=olivedrab10914479]
(axis cs:3577077,0.0999999999999987)
--(axis cs:3577077,0.12);

\path [draw=olivedrab10914479]
(axis cs:35177974,0.829999999999999)
--(axis cs:35177974,0.870000000000005);

\addplot [olivedrab10914479, mark=-, mark size=5, mark options={solid}, only marks, forget plot]
table {
2418 0
5595 0
42417 0
383751 0.01
3577077 0.0999999999999987
35177974 0.829999999999999
};
\addplot [olivedrab10914479, mark=-, mark size=5, mark options={solid}, only marks, forget plot]
table {
2418 0.01
5595 0.01
42417 0.02
383751 0.0300000000000002
3577077 0.12
35177974 0.870000000000005
};
\addplot [ultra thick, goldenrod22917456, mark=diamond*, mark size=3, mark options={solid}]
table {
2418 0
5595 0
42417 0.04
383751 0.249999999999996
3577077 2.62
35177974 19.38
};
\addlegendentry{\systemname-Base}
\path [draw=goldenrod22917456]
(axis cs:2418,0)
--(axis cs:2418,0.01);

\path [draw=goldenrod22917456]
(axis cs:5595,0)
--(axis cs:5595,0.01);

\path [draw=goldenrod22917456]
(axis cs:42417,0.0299999999999998)
--(axis cs:42417,0.0499999999999998);

\path [draw=goldenrod22917456]
(axis cs:383751,0.239999999999998)
--(axis cs:383751,0.2565);

\path [draw=goldenrod22917456]
(axis cs:3577077,2.59999999999999)
--(axis cs:3577077,2.65000000000001);

\path [draw=goldenrod22917456]
(axis cs:35177974,19.2730000000001)
--(axis cs:35177974,19.5270000000001);

\addplot [goldenrod22917456, mark=-, mark size=5, mark options={solid}, only marks, forget plot]
table {
2418 0
5595 0
42417 0.0299999999999998
383751 0.239999999999998
3577077 2.59999999999999
35177974 19.2730000000001
};
\addplot [goldenrod22917456, mark=-, mark size=5, mark options={solid}, only marks, forget plot]
table {
2418 0.01
5595 0.01
42417 0.0499999999999998
383751 0.2565
3577077 2.65000000000001
35177974 19.5270000000001
};
\addplot [ultra thick, tomato2527948, mark=square*, mark size=3, mark options={solid}]
table {
2418 0.0100000000000006
5595 0.0199999999999995
42417 0.0500000000000007
383751 0.409999999999999
3577077 16.965
};
\addlegendentry{RDFTensor}
\path [draw=tomato2527948]
(axis cs:2418,0)
--(axis cs:2418,0.08);

\path [draw=tomato2527948]
(axis cs:5595,0.0099999999999997)
--(axis cs:5595,0.13);

\path [draw=tomato2527948]
(axis cs:42417,0.0399999999999991)
--(axis cs:42417,1.1955);

\path [draw=tomato2527948]
(axis cs:383751,0.320000000000007)
--(axis cs:383751,4.266);

\path [draw=tomato2527948]
(axis cs:3577077,14.3085)
--(axis cs:3577077,26.715);

\addplot [tomato2527948, mark=-, mark size=5, mark options={solid}, only marks, forget plot]
table {
2418 0
5595 0.0099999999999997
42417 0.0399999999999991
383751 0.320000000000007
3577077 14.3085
};
\addplot [tomato2527948, mark=-, mark size=5, mark options={solid}, only marks, forget plot]
table {
2418 0.08
5595 0.13
42417 1.1955
383751 4.266
3577077 26.715
};
\addplot [ultra thick, dodgerblue0143213, mark=*, mark size=3, mark options={solid}]
table {
2418 0.0399999999999991
5595 0.0399999999999991
42417 0.0599999999999987
383751 0.200000000000003
3577077 2.11000000000001
35177974 34.02
};
\addlegendentry{Neo4j}
\path [draw=dodgerblue0143213]
(axis cs:2418,0.0199999999999995)
--(axis cs:2418,0.233);

\path [draw=dodgerblue0143213]
(axis cs:5595,0.0199999999999995)
--(axis cs:5595,0.163);

\path [draw=dodgerblue0143213]
(axis cs:42417,0.0335000000000004)
--(axis cs:42417,0.1965);

\path [draw=dodgerblue0143213]
(axis cs:383751,0.18)
--(axis cs:383751,0.332999999999998);

\path [draw=dodgerblue0143213]
(axis cs:3577077,2.047)
--(axis cs:3577077,3.3655);

\path [draw=dodgerblue0143213]
(axis cs:35177974,30.0760000000002)
--(axis cs:35177974,36.5100000000001);

\addplot [dodgerblue0143213, mark=-, mark size=5, mark options={solid}, only marks, forget plot]
table {
2418 0.0199999999999995
5595 0.0199999999999995
42417 0.0335000000000004
383751 0.18
3577077 2.047
35177974 30.0760000000002
};
\addplot [dodgerblue0143213, mark=-, mark size=5, mark options={solid}, only marks, forget plot]
table {
2418 0.233
5595 0.163
42417 0.1965
383751 0.332999999999998
3577077 3.3655
35177974 36.5100000000001
};
\end{axis}

\end{tikzpicture}
        \caption{CPU time.}
        \label{fig:bsbm_wall_easy}
    \end{subfigure}
    \hfill
    \begin{subfigure}{0.48\textwidth}
        \centering
        \input{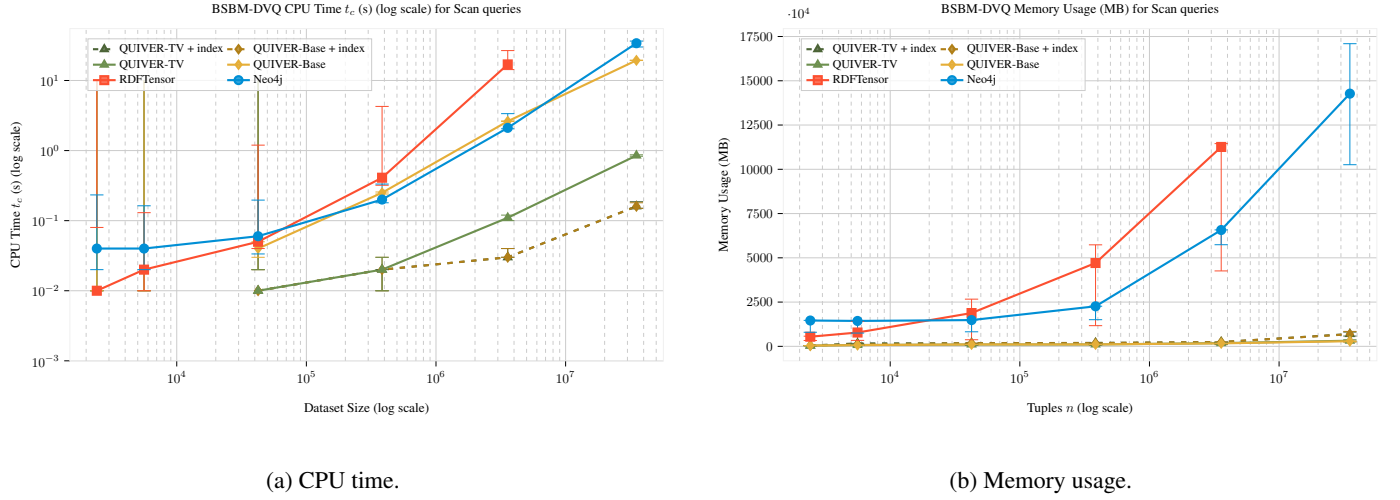}
        \caption{Memory usage.}
        \label{fig:bsbm_mem_easy}
    \end{subfigure}

    \caption{Median CPU time (user + system) (\subref{fig:bsbm_wall_easy}) in seconds and memory usage in megabytes (\subref{fig:bsbm_mem_easy}) for the scan query on the \bsbmname of increasing total tuples $n$. We compare the baseline (RDFTensor~\autocite{Marciniak2025DataTensorsRDF}) and Neo4j against \systemname with and without a vocabulary and with and without indexing. The error bars indicate 5th and 95th percentiles; both x-axes and the CPU time y-axis are logarithmic.}
    \label{fig:bsbm_sys_easy}
\end{figure*}

\section{Experimental Results}
\label{sec:results}
This section reports the empirical evaluation of \systemname. \Cref{sec:bsbmqvq} presents the scalability results on \bsbmname across the six size points of \Cref{tab:bsbm_counts}; \Cref{sec:dbpedia-results} presents the real-world results on \dbpedianame, including the cross-modal join examples. Each subsection compares the four \systemname configurations (\systemname-Base, \systemname-TV, \systemname-Base+Index, \systemname-TV+Index) against the RDFTensor reference implementation on Fuseki, against the two-stage pipeline (\bsbmname only), and against Neo4j with its native vector implementation.

Three headline patterns hold across the results. First, vocabulary-time parsing alone (\systemname-TV) yields a median speedup of up to \bestSpeedupBsbmScanEmbedded on \bsbmname and \bestSpeedupDbpediaScanEmbedded on \dbpedianame over RDFTensor on single-type ranking queries. Second, adding the in-engine \gls{aknn} index (\systemname-TV+Index) reaches up to \bestSpeedupBsbmScanIndex on \bsbmname and \bestSpeedupDbpediaScanIndex on \dbpedianame at the recall--latency operating point; the index additionally renders cross-modal vector joins on \dbpedianame feasible in seconds, whereas every non-indexed configuration exceeds the 30-second timeout on those queries. Third, the asymptotic behavior outlined above, namely a linear scan in $\bigo{\nt}$ and a sub-linear retrieval in $\bigo{\sqrt{\nt}}$, is borne out by the \bsbmname size sweep.

\subsection{\bsbmname Results}\label{sec:bsbmqvq}

\renewcommand\arraystretch{1}
\begin{table*}
\centering
\caption{Speedup of \systemname compared to RDFTensor for our BSBM-DVQ with "Scan" queries. The time is the median time over the group, significant speedups are indicated in \textbf{bold}. Query combinations that timed out are omitted.}
\small
\label{tab:bsbm_speedup_Scan}
\begin{tabular}{lllrrr}
\toprule
 $m$ &  Query Type &  Engine & $t$ (s)  & Speedup  & $p$-value (Baseline)   \\
\midrule
\multirow[c]{8}{*}{$1$} & \multirow[c]{4}{*}{Embedded} & Neo4j & $0.21 \cdot 10^{-1}$ & $0.38$ & $1$ \\
\cline{3-6}
 &  & RDFTensor & $0.76 \cdot 10^{-2}$ & $1$ & $0.50$ \\
\cline{3-6}
 &  & \systemname-Base & $0.23 \cdot 10^{-2}$ & $\bm{3.15}$ & $0.63 \cdot 10^{-4}$ \\
\cline{3-6}
 &  & \systemname-TV & $0.22 \cdot 10^{-2}$ & $\bm{3.42}$ & $0.48 \cdot 10^{-4}$ \\
\cline{2-6} \cline{3-6}
 & \multirow[c]{2}{*}{Index} & \systemname-Base & $0.23 \cdot 10^{-2}$ & $\bm{3.12}$ & $0.70 \cdot 10^{-4}$ \\
\cline{3-6}
 &  & \systemname-TV & $0.24 \cdot 10^{-2}$ & $\bm{3.03}$ & $0.84 \cdot 10^{-4}$ \\
\cline{2-6} \cline{3-6}
 & \multirow[c]{2}{*}{Two Stage} & RDFTensor & $0.50 \cdot 10^{-1}$ & $0.15$ & $1$ \\
\cline{3-6}
 &  & \systemname-Base & $0.15 \cdot 10^{-1}$ & $0.50$ & $1.00$ \\
\cline{1-6} \cline{2-6} \cline{3-6}
\multirow[c]{8}{*}{$10$} & \multirow[c]{4}{*}{Embedded} & Neo4j & $0.22 \cdot 10^{-1}$ & $0.64$ & $0.99$ \\
\cline{3-6}
 &  & RDFTensor & $0.14 \cdot 10^{-1}$ & $1$ & $0.50$ \\
\cline{3-6}
 &  & \systemname-Base & $0.51 \cdot 10^{-2}$ & $\bm{2.55}$ & $0.18 \cdot 10^{-5}$ \\
\cline{3-6}
 &  & \systemname-TV & $0.30 \cdot 10^{-2}$ & $\bm{4.35}$ & $0.28 \cdot 10^{-7}$ \\
\cline{2-6} \cline{3-6}
 & \multirow[c]{2}{*}{Index} & \systemname-Base & $0.32 \cdot 10^{-2}$ & $\bm{4.17}$ & $0.53 \cdot 10^{-7}$ \\
\cline{3-6}
 &  & \systemname-TV & $0.33 \cdot 10^{-2}$ & $\bm{3.96}$ & $0.72 \cdot 10^{-7}$ \\
\cline{2-6} \cline{3-6}
 & \multirow[c]{2}{*}{Two Stage} & RDFTensor & $0.91 \cdot 10^{-1}$ & $0.15$ & $1$ \\
\cline{3-6}
 &  & \systemname-Base & $0.18 \cdot 10^{-1}$ & $0.73$ & $0.76$ \\
\cline{1-6} \cline{2-6} \cline{3-6}
\multirow[c]{8}{*}{$10^{2}$} & \multirow[c]{4}{*}{Embedded} & Neo4j & $0.39 \cdot 10^{-1}$ & $1.27$ & $0.15 \cdot 10^{-3}$ \\
\cline{3-6}
 &  & RDFTensor & $0.52 \cdot 10^{-1}$ & $1$ & $0.50$ \\
\cline{3-6}
 &  & \systemname-Base & $0.30 \cdot 10^{-1}$ & $\bm{1.56}$ & $0.92 \cdot 10^{-9}$ \\
\cline{3-6}
 &  & \systemname-TV & $0.58 \cdot 10^{-2}$ & $\bm{8.15}$ & $0.28 \cdot 10^{-22}$ \\
\cline{2-6} \cline{3-6}
 & \multirow[c]{2}{*}{Index} & \systemname-Base & $0.56 \cdot 10^{-2}$ & $\bm{8.37}$ & $0.22 \cdot 10^{-22}$ \\
\cline{3-6}
 &  & \systemname-TV & $0.57 \cdot 10^{-2}$ & $\bm{8.24}$ & $0.23 \cdot 10^{-22}$ \\
\cline{2-6} \cline{3-6}
 & \multirow[c]{2}{*}{Two Stage} & RDFTensor & $1.92$ & $0.24 \cdot 10^{-1}$ & $1$ \\
\cline{3-6}
 &  & \systemname-Base & $0.13$ & $0.36$ & $1$ \\
\cline{1-6} \cline{2-6} \cline{3-6}
\multirow[c]{7}{*}{$10^{3}$} & \multirow[c]{4}{*}{Embedded} & Neo4j & $0.26$ & $\bm{2.21}$ & $0.41 \cdot 10^{-37}$ \\
\cline{3-6}
 &  & RDFTensor & $0.59$ & $1$ & $0.50$ \\
\cline{3-6}
 &  & \systemname-Base & $0.25$ & $\bm{2.46}$ & $0.49 \cdot 10^{-39}$ \\
\cline{3-6}
 &  & \systemname-TV & $0.14 \cdot 10^{-1}$ & $\bm{4.19 \cdot 10^{1}}$ & $0.36 \cdot 10^{-61}$ \\
\cline{2-6} \cline{3-6}
 & \multirow[c]{2}{*}{Index} & \systemname-Base & $0.83 \cdot 10^{-2}$ & $\bm{7.35 \cdot 10^{1}}$ & $0.12 \cdot 10^{-61}$ \\
\cline{3-6}
 &  & \systemname-TV & $0.81 \cdot 10^{-2}$ & $\bm{7.48 \cdot 10^{1}}$ & $0.11 \cdot 10^{-61}$ \\
\cline{2-6} \cline{3-6}
 & Two Stage & \systemname-Base & $5.95$ & $0.11$ & $1$ \\
\cline{1-6} \cline{2-6} \cline{3-6}
\multirow[c]{6}{*}{$10^{4}$} & \multirow[c]{4}{*}{Embedded} & Neo4j & $1.78$ & $\bm{2.61}$ & $0.20 \cdot 10^{-96}$ \\
\cline{3-6}
 &  & RDFTensor & $4.66$ & $1$ & $0.50$ \\
\cline{3-6}
 &  & \systemname-Base & $2.65$ & $\bm{1.74}$ & $0.38 \cdot 10^{-80}$ \\
\cline{3-6}
 &  & \systemname-TV & $0.13$ & $\bm{3.50 \cdot 10^{1}}$ & $0.57 \cdot 10^{-115}$ \\
\cline{2-6} \cline{3-6}
 & \multirow[c]{2}{*}{Index} & \systemname-Base & $0.13 \cdot 10^{-1}$ & $\bm{3.55 \cdot 10^{2}}$ & $0.22 \cdot 10^{-116}$ \\
\cline{3-6}
 &  & \systemname-TV & $0.19 \cdot 10^{-1}$ & $\bm{2.50 \cdot 10^{2}}$ & $0.32 \cdot 10^{-116}$ \\
\cline{1-6} \cline{2-6} \cline{3-6}
\multirow[c]{6}{*}{$10^{5}$} & \multirow[c]{3}{*}{Embedded} & Neo4j & $1.76 \cdot 10^{1}$ & - & - \\
\cline{3-6}
 &  & \systemname-Base & $1.95 \cdot 10^{1}$ & - & - \\
\cline{3-6}
 &  & \systemname-TV & $0.73$ & - & - \\
\cline{2-6} \cline{3-6}
 & \multirow[c]{2}{*}{Index} & \systemname-Base & $0.63 \cdot 10^{-1}$ & - & - \\
\cline{3-6}
 &  & \systemname-TV & $0.63 \cdot 10^{-1}$ & - & - \\
\cline{2-6} \cline{3-6}
 & Two Stage & RDFTensor & $1.41 \cdot 10^{1}$ & - & - \\
\cline{1-6} \cline{2-6} \cline{3-6}
\bottomrule
\end{tabular}
\end{table*}

\paragraph{Scaling on the scan queries.}

First, we show that \systemname scales beyond small datasets by running the evaluation framework on increasingly large \bsbmname instances on the scan query, as shown in~\Cref{fig:bsbm_timings_easy} . The detailed speedup relative to the RDFTensor implementation and $p$-values are available in \Cref{tab:bsbm_speedup_Scan}. While \systemname-Base without the tensor vocabularies is only slightly faster, the query times still range from \speedupBsbmWorstScanEmbeddedSystemnamebase up to \speedupBsbmBestScanEmbeddedSystemnamebase, with smaller timing margins, yet still significantly outperforming the existing solution. This lead in performance, however, still results in timeouts for larger dataset sizes.

We note that the largest \bsbmname on our RDFTensor times out, leading to incalculable speedups.

Nevertheless, as soon as we enable our vocabularies, we see significant improvements in query performance starting at \worstSpeedupBsbmScanEmbedded, and up to \bestSpeedupBsbmScanEmbedded in median timings where we can attain speedups. Additionally, we extrapolate our measured timings using a second-degree polynomial to predict query values for larger sizes in the graphs. While this extrapolation yields plausible data for the linear systems (i.e., the non-indexed approaches), we exclude it for the indexed version, as it returns implausible negative values. The extrapolations show that the trend of the slower systems is indeed one of rising query time beyond our initial timeout limits.

Moving on to the \gls{aknn}-supported indices, we see even larger speedups, reaching up to \bestSpeedupBsbmScanIndex when using the vocabularies. The speedup is not as drastic for the smaller \bsbmname sizes, starting at just \worstSpeedupBsbmScanIndex and increasing more quickly with size due to the slower timing increase. The speedup for the largest size could not be measured as the baseline timed out. These results are similar for the indexed version without the vocabulary.

The comparison to the two-stage approach, i.e., the pipeline of moving tensors between systems, also yields slow results for the competing systems. Our timing investigation shows query times that are half as fast for the smallest sizes and up to $41 \times$ slower than the baseline on the dataset with $m=10^2$ for the RDFTensor-based version, as listed in \Cref{tab:bsbm_speedup_Scan}.

\begin{figure*}[tb]
    \centering
    \begin{subfigure}{0.48\textwidth}
        \centering
        \input{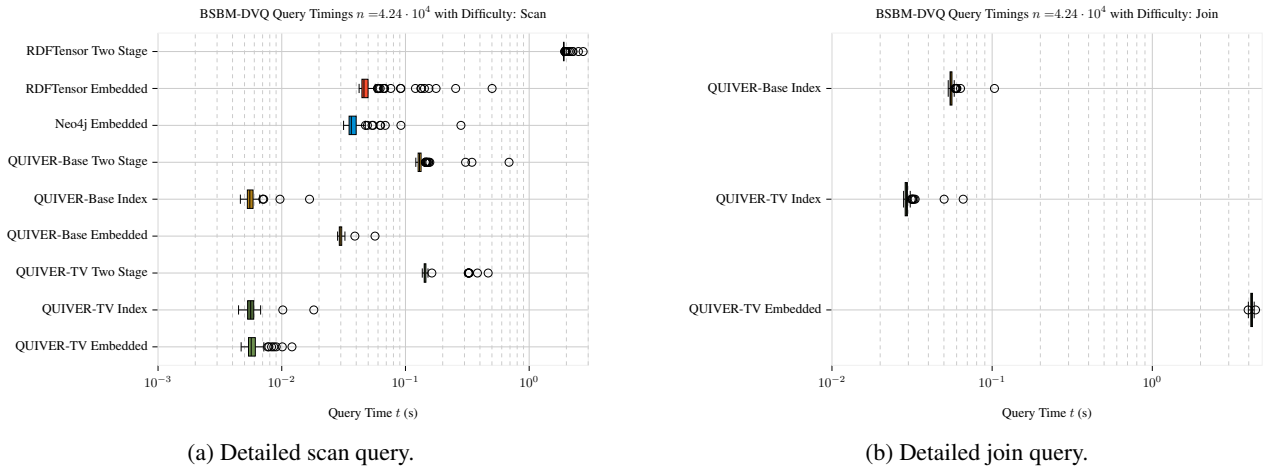}
        \caption{Detailed scan query.}
        \label{fig:bsbm_timings_Scan_100}
    \end{subfigure}
    \hfil
    \begin{subfigure}{0.48\textwidth}
        \centering
        \begin{tikzpicture}[
scale=0.5,
every axis/.style={
                        legend pos=south west, 
                        legend style={
                    font=\small},
                        legend columns=2, 
                        extra x ticks={0.002,0.003,0.004,0.005,0.006,0.007,0.008,0.009000000000000001,0.02,0.03,0.04,0.05,0.06,0.07,0.08,0.09,0.2,0.30000000000000004,0.4,0.5,0.6000000000000001,0.7000000000000001,0.8,0.9,2,3,4,5,6,7,8,9,20,30,40,50,60,70,80,90,200,300,400,500,600,700,800,900},
                        extra x tick labels={ },
                        extra y tick labels={ },
                        extra tick style={
                            tick style={draw=none},
                            major grid style={dashed,red},
                            grid=major,
                        },
                    }
]

\definecolor{darkgoldenrod17612723}{RGB}{176,127,23}
\definecolor{darkolivegreen7610055}{RGB}{76,100,55}
\definecolor{lightgrey203}{RGB}{203,203,203}
\definecolor{olivedrab10914479}{RGB}{109,144,79}
\definecolor{whitesmoke240}{RGB}{240,240,240}

\begin{axis}[
axis line style={whitesmoke240},
height=1.2\linewidth,
log basis x={10},
tick align=outside,
tick pos=left,
title={ BSBM-DVQ Query Timings \(\displaystyle n=\)\(\displaystyle 4.24 \cdot 10^{4}\) with Difficulty: Join},
width=1.5\linewidth,
x grid style={lightgrey203},
xlabel={Query Time \(\displaystyle t\) (s)},
xmajorgrids,
xmin=0.01, xmax=4.85256218711147,
xmode=log,
xtick style={color=black},
xtick={0.001,0.01,0.1,1,10,100},
xticklabels={
  \(\displaystyle {10^{-3}}\),
  \(\displaystyle {10^{-2}}\),
  \(\displaystyle {10^{-1}}\),
  \(\displaystyle {10^{0}}\),
  \(\displaystyle {10^{1}}\),
  \(\displaystyle {10^{2}}\)
},
y grid style={lightgrey203},
ymajorgrids,
ymin=0.5, ymax=3.5,
ytick style={color=black},
ytick={1,2,3},
yticklabels={\systemname-TV Embedded,\systemname-TV Index,\systemname-Base Index}
]
\path [draw=black, fill=olivedrab10914479]
(axis cs:4.12180974869989,0.85)
--(axis cs:4.12180974869989,1.15)
--(axis cs:4.21747839575983,1.15)
--(axis cs:4.21747839575983,0.85)
--(axis cs:4.12180974869989,0.85)
--cycle;
\addplot [black]
table {
4.12180974869989 1
3.97897267597727 1
};
\addplot [black]
table {
4.21747839575983 1
4.32670947292354 1
};
\addplot [black]
table {
3.97897267597727 0.925
3.97897267597727 1.075
};
\addplot [black]
table {
4.32670947292354 0.925
4.32670947292354 1.075
};
\addplot [black, mark=o, mark size=3, mark options={solid,fill opacity=0}, only marks]
table {
3.96505898004398 1
4.41142017010134 1
};
\path [draw=black, fill=darkolivegreen7610055]
(axis cs:0.0286461529321968,1.85)
--(axis cs:0.0286461529321968,2.15)
--(axis cs:0.0295673559012357,2.15)
--(axis cs:0.0295673559012357,1.85)
--(axis cs:0.0286461529321968,1.85)
--cycle;
\addplot [black]
table {
0.0286461529321968 2
0.0280060400255024 2
};
\addplot [black]
table {
0.0295673559012357 2
0.0308103520656004 2
};
\addplot [black]
table {
0.0280060400255024 1.925
0.0280060400255024 2.075
};
\addplot [black]
table {
0.0308103520656004 1.925
0.0308103520656004 2.075
};
\addplot [black, mark=o, mark size=3, mark options={solid,fill opacity=0}, only marks]
table {
0.0501506680157035 2
0.0330636600265279 2
0.0325217109639197 2
0.0312748930882662 2
0.0659579550847411 2
0.0319608770078048 2
0.0317108579911291 2
0.0318433409556746 2
};
\path [draw=black, fill=darkgoldenrod17612723]
(axis cs:0.0546071594872046,2.85)
--(axis cs:0.0546071594872046,3.15)
--(axis cs:0.0561299040564336,3.15)
--(axis cs:0.0561299040564336,2.85)
--(axis cs:0.0546071594872046,2.85)
--cycle;
\addplot [black]
table {
0.0546071594872046 3
0.053197614965029 3
};
\addplot [black]
table {
0.0561299040564336 3
0.0579926029313355 3
};
\addplot [black]
table {
0.053197614965029 2.925
0.053197614965029 3.075
};
\addplot [black]
table {
0.0579926029313355 2.925
0.0579926029313355 3.075
};
\addplot [black, mark=o, mark size=3, mark options={solid,fill opacity=0}, only marks]
table {
0.103429469978437 3
0.0585168780526146 3
0.0593662150204181 3
0.0634789150208234 3
0.0602613299852237 3
0.0595048000104725 3
0.059948330046609 3
0.0588485529879108 3
};
\addplot [semithick, black]
table {
4.16664859751472 0.85
4.16664859751472 1.15
};
\addplot [semithick, black]
table {
0.0291500714956782 1.85
0.0291500714956782 2.15
};
\addplot [semithick, black]
table {
0.0551398025127127 2.85
0.0551398025127127 3.15
};
\end{axis}

\end{tikzpicture}
        \caption{Detailed join query.}
        \label{fig:bsbm_timings_Join_100}
    \end{subfigure}
    \caption{Box plot for the scan (\subref{fig:bsbm_timings_Scan_100}) and join (\subref{fig:bsbm_timings_Scan_100}) query on \bsbmname $n=4.24 \cdot 10^{4}$. We compare the baseline (RDFTensor~\autocite{Marciniak2025DataTensorsRDF}) and Neo4j against our \systemname with and without a vocabulary and with and without indexing. We note that the time $t$ is logarithmic.}
    \label{fig:bsbm_timings_boxplots}
\end{figure*}

We therefore do not investigate them beyond this stage and difficulty, as they either time out sporadically for larger $m$, proving them to be infeasible for any complex \gls{kg} within reasonable time spans, or become increasingly inaccurate due to the limited transfer of data.
The \gls{pg} version in Neo4j performs similarly to the other base implementations, outperforming the Fuseki-based baseline up to a factor of \speedupBsbmBestScanEmbeddedNeoj at larger sizes, with no indexed queries performed, as they cannot be expressed in the system as noted above.Further timing distributions supporting our results are detailed in \Cref{fig:bsbm_timings_Scan_100}. \systemname shows a strong lead, with very tight distributions around the medians compared to the other implementations. The two-stage result is naturally slower, as it uses the same pipeline for all systems.
The hard queries in \Cref{fig:bsbm_timings_Scan_100} show that the base and tensor vocabulary indices also have a tight distribution, with just one outlier -- the initial index build.

The investigation into query times on the \gls{aknn} indices is enriched by sweeping over a range of \texttt{searchK} values for both the \gls{hnsw} and \gls{ivf} \gls{aknn} implementations. Through this sweep, the efficiency/effectiveness trade-off of the indices introduced in \Cref{sec:optim} is investigated. This evaluation tracks the warm runtime, i.e., time after running the query once, over the recall\@16 and is shown in \Cref{fig:bsbm_timings_index}. The figure shows a general trend: both versions are faster when \texttt{searchK} is lower; in turn, increasing $ef_{\mathrm{search}}$ and $K_{\mathrm{IVF}}$ lead to  better query performance.
The recall is computed by comparing to the \texttt{naive} implementation, i.e., a full scan over both sides. These findings are in line with \gls{aknn} evaluations on other systems~\autocite{Douze2025FaissLibrary}, suggesting that users also have to make an efficiency/effectiveness trade-off with our system.

All query executions by \systemname carried out in our evaluation campaign, additionally, require much fewer resources than the existing implementation, as our implementation consistently maintains a lower memory footprint of below 500 megabytes for the non-indexed version, as shown in \Cref{fig:bsbm_mem_easy}. The indexed version requires slightly more memory, rising from 45 megabytes to 843 megabytes. RDFTensor, however, requires at least 543 megabytes of memory already for the smallest configuration, increasing linearly to more than 10 gigabytes for $m=10^4$.
Neo4j similarly requires a comparatively more memory, starting at 1458 megabytes and rising to 14.3 gigabytes, with peaks of 17.1 gigabytes for the largest version of \bsbmname.
These resource usage results also indicate strong efficiency gains in our system variations, as the memory usage shows a strong lead of \systemname across configurations. This can be attributed to the fact that the compared systems are Java-based, requiring comparatively large amounts of memory just for the Java Virtual Machine, whereas QLever, our base, uses an efficient C++ implementation.
Only the indices for the largest $m$ start to require more memory within our implementation, whereas the other systems, especially Neo4j, rise.

The CPU time $t_m$ in \Cref{fig:bsbm_wall_easy} behaves similarly to the query times, with the Fuseki-based implementation requiring comparatively more CPU time over all tested queries. Note that the CPU times below 10 ms are cut due to measurement limitations.

\renewcommand\arraystretch{1.2}
\begin{table*}
\centering
\caption{Speedup of \systemname compared to RDFTensor for our BSBM-DVQ with "Join" queries. The time is the median time over the group, significant speedups are indicated in \textbf{bold}. Query combinations that timed out are omitted.}
\small
\label{tab:bsbm_speedup_Join}
\begin{tabular}{lllrrr}
\toprule
 $m$ &  Query Type &  Engine & $t$ (s)  & Speedup  & $p$-value (Baseline)   \\
\midrule
\multirow[c]{6}{*}{$1$} & \multirow[c]{4}{*}{Embedded} & Neo4j & $0.26 \cdot 10^{-1}$ & $0.47$ & $1.00$ \\
\cline{3-6}
 &  & RDFTensor & $0.13 \cdot 10^{-1}$ & $1$ & $0.50$ \\
\cline{3-6}
 &  & \systemname-Base & $0.94 \cdot 10^{-2}$ & $1.22$ & $0.83 \cdot 10^{-2}$ \\
\cline{3-6}
 &  & \systemname-TV & $0.28 \cdot 10^{-2}$ & $\bm{4.11}$ & $0.75 \cdot 10^{-7}$ \\
\cline{2-6} \cline{3-6}
 & \multirow[c]{2}{*}{Index} & \systemname-Base & $0.38 \cdot 10^{-2}$ & $\bm{3.12}$ & $0.89 \cdot 10^{-6}$ \\
\cline{3-6}
 &  & \systemname-TV & $0.36 \cdot 10^{-2}$ & $\bm{3.13}$ & $0.45 \cdot 10^{-6}$ \\
\cline{1-6} \cline{2-6} \cline{3-6}
\multirow[c]{6}{*}{$10$} & \multirow[c]{4}{*}{Embedded} & Neo4j & $0.36$ & $\bm{2.46}$ & $0.14 \cdot 10^{-84}$ \\
\cline{3-6}
 &  & RDFTensor & $0.84$ & $1$ & $0.50$ \\
\cline{3-6}
 &  & \systemname-Base & $0.99$ & $0.89$ & $1$ \\
\cline{3-6}
 &  & \systemname-TV & $0.34 \cdot 10^{-1}$ & $\bm{2.58 \cdot 10^{1}}$ & $0.34 \cdot 10^{-106}$ \\
\cline{2-6} \cline{3-6}
 & \multirow[c]{2}{*}{Index} & \systemname-Base & $0.77 \cdot 10^{-2}$ & $\bm{1.15 \cdot 10^{2}}$ & $0.75 \cdot 10^{-108}$ \\
\cline{3-6}
 &  & \systemname-TV & $0.47 \cdot 10^{-2}$ & $\bm{1.88 \cdot 10^{2}}$ & $0.49 \cdot 10^{-108}$ \\
\cline{1-6} \cline{2-6} \cline{3-6}
\multirow[c]{3}{*}{$10^{2}$} & Embedded & \systemname-TV & $4.17$ & - & - \\
\cline{2-6} \cline{3-6}
 & \multirow[c]{2}{*}{Index} & \systemname-Base & $0.56 \cdot 10^{-1}$ & - & - \\
\cline{3-6}
 &  & \systemname-TV & $0.30 \cdot 10^{-1}$ & - & - \\
\cline{1-6} \cline{2-6} \cline{3-6}
\multirow[c]{2}{*}{$10^{3}$} & \multirow[c]{2}{*}{Index} & \systemname-Base & $0.77$ & - & - \\
\cline{3-6}
 &  & \systemname-TV & $0.53$ & - & - \\
\cline{1-6} \cline{2-6} \cline{3-6}
\multirow[c]{2}{*}{$10^{4}$} & \multirow[c]{2}{*}{Index} & \systemname-Base & $2.85$ & - & - \\
\cline{3-6}
 &  & \systemname-TV & $2.73$ & - & - \\
\cline{1-6} \cline{2-6} \cline{3-6}
\bottomrule
\end{tabular}
\end{table*}

\paragraph{Translating to semantic joins.}
The join queries for our scalability tests on \cgls{bsbm} show similar trends in \Cref{fig:bsbm_timings_hard} and \Cref{tab:bsbm_speedup_Join}, with a few notable exceptions. First, we achieve no significant speedup for \systemname-Base. The other variants, though, perform significantly faster for the smallest sizes. Additionally, we see significant improvements for \systemname-TV over the Fuseki-based baseline across all benchmark sets that could run it, ranging from \worstSpeedupBsbmJoinEmbedded to \bestSpeedupBsbmJoinEmbedded. Our index-based solution also significantly outperforms the baseline, achieving a speedup of up to \bestSpeedupBsbmJoinIndex and running for all but the largest size. We cannot, however, calculate the speedup beyond $m=10$, since the baseline failed to finish within our timeout window, as does \systemname-Base.

This attests to the efficiency of \systemname, as we can operate over a much larger \gls{kg} than any other compared system using our \gls{aknn} integration.

\paragraph{Investigating Sub-Timings}
Analysing the timings of the individual calls enables us to gain insights into our optimisations, where we can clearly dissect each optimisation level by concrete timing improvements. The sub-timings themselves are provided in \Cref{tab:bsbm_subtimings_Scan_1000,tab:bsbm_subtimings_Join_10}. The scan queries of \Cref{tab:bsbm_subtimings_Scan_1000} provide these insights directly, where the engine-level integration ($a.$) already greatly reduces time spent on the actual dot-product computation and JSON parsing. The move to the tensor vocabulary ($b.$) in \systemname-TV enhances this even further, almost eliminating the time spent on parsing, both the JSON input from the query and the tensor buffer.

Our final optimisation, the integration of an ($c.$) \gls{aknn}, pushes the query times to be even smaller

These sub-timing insights are also applicable to the join query in \Cref{tab:bsbm_subtimings_Join_10}, where the engine speed and vocabulary improvements also shift the timings from the dot product and parsing towards non-reducible tasks, like the disk read.
The index, more prominently, however, reduces the query time even further due to shifting the complete cross-product to a single index build, which itself is comparatively fast, and repeated diminishing lookup times compared to the complete $\bigo{n^2}$ join.

\begin{table*}
\centering
\caption{Sub-Timings for the \bsbmname dataset with query "Scan" and $m=$$10^{3}$. The timings are for a cold-start query after suming over the timing hits in milliseconds, with the number hits $h$ in parentheses.}
\tiny
\label{tab:bsbm_subtimings_Scan_1000}
\begin{tabular}{llllllllll}
\toprule
 Engine &  Query Type & Dot Product  & Index Creation  & Index Lookup  & JSON Parse  & RDF Read  & TV Parse  & TV Read  & Total   \\
\midrule
RDFTensor & Embedded & 420.09 (20548) & - & - & 453.10 (1000) & - & - & - & 1149.48 (1) \\
\cline{1-10} \cline{2-10}
\multirow[c]{2}{*}{\systemname-Base} & Embedded & 2.52 (20548) & - & - & 5.92 (20551) & 115.35 (20548) & - & - & 398.29 (1) \\
\cline{2-10}
 & Index & - & 32.68 (1) & 0.02 (1) & 0.33 (1001) & 6.07 (1001) & - & - & 41.18 (1) \\
\cline{1-10} \cline{2-10}
\multirow[c]{2}{*}{\systemname-TV} & Embedded & 2.64 (20548) & - & - & 0.01 (3) & - & 0.67 (20548) & 34.36 (20548) & 108.41 (1) \\
\cline{2-10}
 & Index & - & 19.14 (1) & 0.02 (1) & 0.00 (1) & 0.00 (1) & 0.03 (1000) & 2.32 (1000) & 28.06 (1) \\
\cline{1-10} \cline{2-10}
\bottomrule
\end{tabular}
\end{table*}

\begin{table*}
\centering
\caption{Sub-Timings for the \bsbmname dataset with query "Join" and $m=$$10$. The timings are for a cold-start query after suming over the timing hits in milliseconds, with the number hits $h$ in parentheses.}
\tiny
\label{tab:bsbm_subtimings_Join_10}
\begin{tabular}{llllllllll}
\toprule
 Engine &  Query Type & Dot Product  & Index Creation  & Index Lookup  & JSON Parse  & RDF Read  & TV Parse  & TV Read  & Total   \\
\midrule
RDFTensor & Embedded & 667.90 (45369) & - & - & 259.46 (144) & - & - & - & 1529.74 (1) \\
\cline{1-10} \cline{2-10}
\multirow[c]{2}{*}{\systemname-Base} & Embedded & 5.96 (45369) & - & - & 26.63 (90738) & 386.79 (90738) & - & - & 1483.02 (1) \\
\cline{2-10}
 & Index & - & 16.54 (1) & 0.93 (213) & 0.14 (426) & 2.14 (426) & - & - & 31.45 (1) \\
\cline{1-10} \cline{2-10}
\multirow[c]{2}{*}{\systemname-TV} & Embedded & 5.84 (45369) & - & - & - & - & 2.81 (90738) & 149.95 (90738) & 366.47 (1) \\
\cline{2-10}
 & Index & - & 11.63 (1) & 1.20 (213) & - & - & 0.02 (426) & 1.11 (426) & 23.59 (1) \\
\cline{1-10} \cline{2-10}
\bottomrule
\end{tabular}
\end{table*}

\setlength\tabcolsep{0.3em}

\begin{table*}
\centering
\caption{Speedup of \systemname compared to RDFTensor for our DBpedia-MS. The time is the median time over the group, significant speedups are indicated in \textbf{bold}. Query combinations that timed out are omitted.}
\small
\label{tab:dbpedia_speedup}
\begin{tabular}{lllrrr}
\toprule
 Difficulty &  Query Type &  Engine & $t$ (s)  & Speedup  & $p$-value (Baseline)   \\
\midrule
\multirow[c]{2}{*}{Join} & \multirow[c]{2}{*}{Index} & \systemname-Base & $1.20$ & - & - \\
\cline{3-6}
 &  & \systemname-TV & $0.72$ & - & - \\
\cline{1-6} \cline{2-6} \cline{3-6}
\multirow[c]{5}{*}{Scan} & \multirow[c]{3}{*}{Embedded} & RDFTensor & $1.13$ & $1$ & $0.50$ \\
\cline{3-6}
 &  & \systemname-Base & $0.75$ & $\bm{1.52}$ & $0.41 \cdot 10^{-12}$ \\
\cline{3-6}
 &  & \systemname-TV & $0.57 \cdot 10^{-1}$ & $\bm{2.00 \cdot 10^{1}}$ & $0.16 \cdot 10^{-42}$ \\
\cline{2-6} \cline{3-6}
 & \multirow[c]{2}{*}{Index} & \systemname-Base & $0.13 \cdot 10^{-1}$ & $\bm{9.27 \cdot 10^{1}}$ & $0.23 \cdot 10^{-44}$ \\
\cline{3-6}
 &  & \systemname-TV & $0.12 \cdot 10^{-1}$ & $\bm{9.78 \cdot 10^{1}}$ & $0.33 \cdot 10^{-44}$ \\
\cline{1-6} \cline{2-6} \cline{3-6}
\bottomrule
\end{tabular}
\end{table*}

\subsection{\dbpedianame Results}\label{sec:dbpedia-results}

The DBpedia use case demonstrates the practical benefits of employing such a vector-based search system in combination with a \gls{kg}, showing, again, significant performance improvements with a tensor vocabulary, while also highlighting the practical limitations of indexed approaches. These results are further supported by the demonstration of multimodal join results, showing novel applications enabled by our solution.

\begin{figure*}[tb]
    \centering
    \begin{subfigure}{0.48\textwidth}

        \begin{tikzpicture}[
scale=0.5,
every axis/.style={
                        legend pos=south west, 
                        legend style={
                    font=\small},
                        legend columns=2, 
                        extra x ticks={0.02,0.03,0.04,0.05,0.06,0.07,0.08,0.09,0.2,0.30000000000000004,0.4,0.5,0.6000000000000001,0.7000000000000001,0.8,0.9,2,3,4,5,6,7,8,9,20,30,40,50,60,70,80,90},
                        extra x tick labels={ },
                        extra y tick labels={ },
                        extra tick style={
                            tick style={draw=none},
                            major grid style={dashed,red},
                            grid=major,
                        },
                    }
]

\definecolor{darkgoldenrod17612723}{RGB}{176,127,23}
\definecolor{dodgerblue0143213}{RGB}{0,143,213}
\definecolor{firebrick207332}{RGB}{207,33,2}
\definecolor{goldenrod22917456}{RGB}{229,174,56}
\definecolor{lightgrey203}{RGB}{203,203,203}
\definecolor{tomato2527948}{RGB}{252,79,48}
\definecolor{whitesmoke240}{RGB}{240,240,240}

\begin{axis}[
axis line style={whitesmoke240},
height=1.2\linewidth,
log basis x={10},
tick align=outside,
tick pos=left,
title={ DBpedia-MS Query Timings \(\displaystyle n=\)\(\displaystyle 4.22 \cdot 10^{8}\) with Difficulty: Scan},
width=1.5\linewidth,
x grid style={lightgrey203},
xlabel={Query Time \(\displaystyle t\) (s)},
xmajorgrids,
xmin=0.01, xmax=8.48023181826575,
xmode=log,
xtick style={color=black},
xtick={0.001,0.01,0.1,1,10,100},
xticklabels={
  \(\displaystyle {10^{-3}}\),
  \(\displaystyle {10^{-2}}\),
  \(\displaystyle {10^{-1}}\),
  \(\displaystyle {10^{0}}\),
  \(\displaystyle {10^{1}}\),
  \(\displaystyle {10^{2}}\)
},
y grid style={lightgrey203},
ymajorgrids,
ymin=0.5, ymax=5.5,
ytick style={color=black},
ytick={1,2,3,4,5},
yticklabels={
  \systemname-TV Index,
  \systemname-TV Embedded,
  \systemname-Base Index,
  \systemname-Base Embedded,
  RDFTensor Embedded
}
]
\path [draw=black, fill=darkgoldenrod17612723]
(axis cs:0.0111426599905826,0.75)
--(axis cs:0.0111426599905826,1.25)
--(axis cs:0.0120684129942673,1.25)
--(axis cs:0.0120684129942673,0.75)
--(axis cs:0.0111426599905826,0.75)
--cycle;
\addplot [black]
table {
0.0111426599905826 1
0.0100049750180915 1
};
\addplot [black]
table {
0.0120684129942673 1
0.0133332420373335 1
};
\addplot [black]
table {
0.0100049750180915 0.875
0.0100049750180915 1.125
};
\addplot [black]
table {
0.0133332420373335 0.875
0.0133332420373335 1.125
};
\addplot [black, mark=o, mark size=3, mark options={solid,fill opacity=0}, only marks]
table {
0.362075085984543 1
0.0143740959465503 1
0.0134827479487285 1
};
\path [draw=black, fill=goldenrod22917456]
(axis cs:0.0561874317063484,1.75)
--(axis cs:0.0561874317063484,2.25)
--(axis cs:0.0578203975455835,2.25)
--(axis cs:0.0578203975455835,1.75)
--(axis cs:0.0561874317063484,1.75)
--cycle;
\addplot [black]
table {
0.0561874317063484 2
0.0541862789541482 2
};
\addplot [black]
table {
0.0578203975455835 2
0.059850709978491 2
};
\addplot [black]
table {
0.0541862789541482 1.875
0.0541862789541482 2.125
};
\addplot [black]
table {
0.059850709978491 1.875
0.059850709978491 2.125
};
\addplot [black, mark=o, mark size=3, mark options={solid,fill opacity=0}, only marks]
table {
0.0695566210197284 2
0.0604392959503456 2
0.0650925899390131 2
0.0606688560219481 2
0.0620710829971358 2
};
\path [draw=black, fill=firebrick207332]
(axis cs:0.01171245452133,2.75)
--(axis cs:0.01171245452133,3.25)
--(axis cs:0.0130452988087199,3.25)
--(axis cs:0.0130452988087199,2.75)
--(axis cs:0.01171245452133,2.75)
--cycle;
\addplot [black]
table {
0.01171245452133 3
0.010607860982418 3
};
\addplot [black]
table {
0.0130452988087199 3
0.0149000739911571 3
};
\addplot [black]
table {
0.010607860982418 2.875
0.010607860982418 3.125
};
\addplot [black]
table {
0.0149000739911571 2.875
0.0149000739911571 3.125
};
\addplot [black, mark=o, mark size=3, mark options={solid,fill opacity=0}, only marks]
table {
1.02959958801512 3
0.016090334043838 3
0.0170608029002323 3
0.0168833940988406 3
0.0156682449160143 3
0.0167084279237315 3
};
\path [draw=black, fill=tomato2527948]
(axis cs:0.74404370947741,3.75)
--(axis cs:0.74404370947741,4.25)
--(axis cs:0.75477375232731,4.25)
--(axis cs:0.75477375232731,3.75)
--(axis cs:0.74404370947741,3.75)
--cycle;
\addplot [black]
table {
0.74404370947741 4
0.73925457696896 4
};
\addplot [black]
table {
0.75477375232731 4
0.770056243985891 4
};
\addplot [black]
table {
0.73925457696896 3.875
0.73925457696896 4.125
};
\addplot [black]
table {
0.770056243985891 3.875
0.770056243985891 4.125
};
\addplot [black, mark=o, mark size=3, mark options={solid,fill opacity=0}, only marks]
table {
0.772210821975023 4
0.791109845973551 4
0.771438262076117 4
0.771283105015755 4
};
\path [draw=black, fill=dodgerblue0143213]
(axis cs:1.06366570500541,4.75)
--(axis cs:1.06366570500541,5.25)
--(axis cs:1.21159057374462,5.25)
--(axis cs:1.21159057374462,4.75)
--(axis cs:1.06366570500541,4.75)
--cycle;
\addplot [black]
table {
1.06366570500541 5
0.853818234987557 5
};
\addplot [black]
table {
1.21159057374462 5
1.39869092905428 5
};
\addplot [black]
table {
0.853818234987557 4.875
0.853818234987557 5.125
};
\addplot [black]
table {
1.39869092905428 4.875
1.39869092905428 5.125
};
\addplot [black, mark=o, mark size=3, mark options={solid,fill opacity=0}, only marks]
table {
7.70930165296886 5
1.58572288299911 5
};
\addplot [semithick, black]
table {
0.0116351029719226 0.75
0.0116351029719226 1.25
};
\addplot [semithick, black]
table {
0.0570185990072787 1.75
0.0570185990072787 2.25
};
\addplot [semithick, black]
table {
0.012275215005502 2.75
0.012275215005502 3.25
};
\addplot [semithick, black]
table {
0.748441236501094 3.75
0.748441236501094 4.25
};
\addplot [semithick, black]
table {
1.13756532152183 4.75
1.13756532152183 5.25
};
\end{axis}

\end{tikzpicture}
        \caption{Scan query.}
        \label{fig:dbpedia_box_Scan}
    \end{subfigure}
    \hfill
    \begin{subfigure}{0.48\textwidth}

        \begin{tikzpicture}[
scale=0.5,
every axis/.style={
                        legend pos=south west, 
                        legend style={
                    font=\small},
                        legend columns=2, 
                        extra x ticks={0.02,0.03,0.04,0.05,0.06,0.07,0.08,0.09,0.2,0.30000000000000004,0.4,0.5,0.6000000000000001,0.7000000000000001,0.8,0.9,2,3,4,5,6,7,8,9,20,30,40,50,60,70,80,90},
                        extra x tick labels={ },
                        extra y tick labels={ },
                        extra tick style={
                            tick style={draw=none},
                            major grid style={dashed,red},
                            grid=major,
                        },
                    }
]

\definecolor{darkgoldenrod17612723}{RGB}{176,127,23}
\definecolor{firebrick207332}{RGB}{207,33,2}
\definecolor{lightgrey203}{RGB}{203,203,203}
\definecolor{whitesmoke240}{RGB}{240,240,240}

\begin{axis}[
axis line style={whitesmoke240},
height=1.2\linewidth,
log basis x={10},
tick align=outside,
tick pos=left,
title={ DBpedia-MS Query Timings \(\displaystyle n=\)\(\displaystyle 4.22 \cdot 10^{8}\) with Difficulty: Join},
width=1.5\linewidth,
x grid style={lightgrey203},
xlabel={Query Time \(\displaystyle t\) (s)},
xmajorgrids,
xmin=0.1, xmax=2.65555522768991,
xmode=log,
xtick style={color=black},
xtick={0.01,0.1,1,10,100},
xticklabels={
  \(\displaystyle {10^{-2}}\),
  \(\displaystyle {10^{-1}}\),
  \(\displaystyle {10^{0}}\),
  \(\displaystyle {10^{1}}\),
  \(\displaystyle {10^{2}}\)
},
y grid style={lightgrey203},
ymajorgrids,
ymin=0.5, ymax=2.5,
ytick style={color=black},
ytick={1,2},
yticklabels={\systemname-TV Index,\systemname-Base Index}
]
\path [draw=black, fill=darkgoldenrod17612723]
(axis cs:0.712600114522502,0.925)
--(axis cs:0.712600114522502,1.075)
--(axis cs:0.716594244062435,1.075)
--(axis cs:0.716594244062435,0.925)
--(axis cs:0.712600114522502,0.925)
--cycle;
\addplot [black]
table {
0.712600114522502 1
0.708236959995702 1
};
\addplot [black]
table {
0.716594244062435 1
0.722583761089481 1
};
\addplot [black]
table {
0.708236959995702 0.9625
0.708236959995702 1.0375
};
\addplot [black]
table {
0.722583761089481 0.9625
0.722583761089481 1.0375
};
\addplot [black, mark=o, mark size=3, mark options={solid,fill opacity=0}, only marks]
table {
1.26978110498749 1
0.728096842998639 1
0.723063430981711 1
};
\path [draw=black, fill=firebrick207332]
(axis cs:1.18397379093221,1.925)
--(axis cs:1.18397379093221,2.075)
--(axis cs:1.19473727431614,2.075)
--(axis cs:1.19473727431614,1.925)
--(axis cs:1.18397379093221,1.925)
--cycle;
\addplot [black]
table {
1.18397379093221 2
1.17787714092992 2
};
\addplot [black]
table {
1.19473727431614 2
1.21065268805251 2
};
\addplot [black]
table {
1.17787714092992 1.9625
1.17787714092992 2.0375
};
\addplot [black]
table {
1.21065268805251 1.9625
1.21065268805251 2.0375
};
\addplot [black, mark=o, mark size=3, mark options={solid,fill opacity=0}, only marks]
table {
2.41414111608174 2
1.21391309506726 2
1.23683749896009 2
1.22173694206867 2
1.21778416994493 2
1.24801918002777 2
1.22267953702249 2
1.23808185302187 2
1.21295360499062 2
1.22069651796483 2
1.21405662898906 2
1.22336596203968 2
};
\addplot [semithick, black]
table {
0.71428938297322 0.925
0.71428938297322 1.075
};
\addplot [semithick, black]
table {
1.18769426649669 1.925
1.18769426649669 2.075
};
\end{axis}

\end{tikzpicture}
        \caption{Join query.}
        \label{fig:dbpedia_box_Join}
    \end{subfigure}
    \caption{Box plot for the scan (\subref{fig:dbpedia_box_Scan}) and join (\subref{fig:dbpedia_box_Join}) queries on DBpedia~\autocite{Lehmann2015DBpediaA}. We compare the baseline (RDFTensor~\autocite{Marciniak2025DataTensorsRDF}) against our \systemname with and without a vocabulary and with and without indexing. We note that the time $t$ is logarithmic.}
    \label{fig:dbpedia_timings}
\end{figure*}
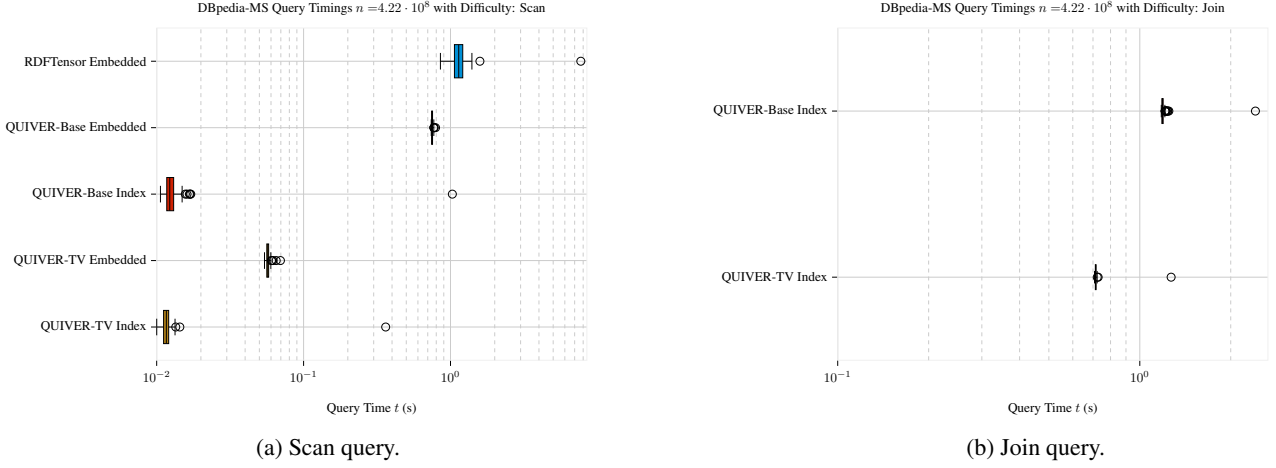

\paragraph{Timings on a real-world \gls{kg}.}
The execution times of the sample queries on DBpedia show similar trends, with some results illustrating practical implications. The results indicate a significant speedup for all versions on the scan single-type ranking query, ranging from \speedupDbpediaBestScanEmbeddedSystemnamebase of the base version, to \speedupDbpediaBestScanEmbeddedSystemnametv with the tensor vocabularies, and finally up to \bestSpeedupDbpediaScanIndex for the indexed version. These were the only ones we were able to compare, since no non-indexed version completed the join query in time. A more detailed overview of the timings is also provided in~\Cref{tab:dbpedia_speedup}.The timing box plots in \Cref{fig:dbpedia_timings} give further insights into these speedups, showing that a single outlier query is comparatively slow in the indexed query executions, with the others being quite dense. This artifact is due to the index build requiring an initially longer execution to build the query, where the non-tensor vocabulary version includes the parsing, slowing it down slightly -- in line with our sub-timing results above.

The join query timings, also shown in~\Cref{tab:dbpedia_speedup}, however, indicate that an index is especially useful for semantic joins over complex data, like the executed query, as introduced in~\Cref{fig:query-possibilities}. The timeout of the non-indexed approaches demonstrates that the indexed version, while incurring some initial overhead, as observed for the scan queries, enables the more complex use case of joins over large datasets.

\tikzexternaldisable
\begin{table*}[tb]

    \newcommand{\imgheight}{2em}
    \newcommand{\imgwidth}{4em}
    \newcommand{\tablespacing}{0em}
    \newcommand{\padfig}[1]{\begin{tikzpicture}
            \node (p) [inner sep=1pt, text height=\imgheight] {#1};
        \end{tikzpicture}
    }
    \newcommand{\resulttxt}[2][0.3\linewidth]{
        \parbox{#1}{\vspace{-1.75em} \raggedleft \texttt{#2}}
    }
    \newcommand{\tdots}[1][0.3\linewidth]{
        \parbox{#1}{\vspace{0.5em} \raggedleft \ldots}
    }
    \newcommand{\resultdist}[2][0.05\linewidth]{
        \parbox{#1}{\vspace{-1.75em} \raggedright \texttt{#2}}
    }
    \centering

    \caption{Result overview for the join query on DBpedia with Wikimedia Thumbnails on the top $k=5$ results. We observe that the most similar-looking villages have remote or scenic thumbnails. We emphasise that we do not evaluate the capabilities of the model, but show that such a query is feasible \emph{at all}.}
    \renewcommand\arraystretch{0}
    \setlength\tabcolsep{0.2em}
    \small
    \label{tab:dbpedia_results_hard}
    \begin{tabular}{rrccl}
    \toprule
    NaturalPlace                               & Village                          & Image NaturalPlace                                                                                                                   & Image Village                                                                                                                         & Distance            \\
    \midrule
    \resulttxt{dbr:Ngang\_Pass}                & \resulttxt{dbr:Hathlangoo}       & \padfig{\includegraphics[height=\imgheight, width=\imgwidth, keepaspectratio]{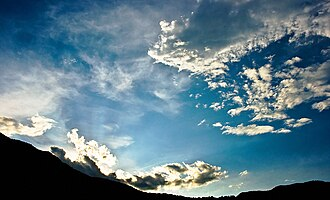}} & \padfig{\includegraphics[height=\imgheight, width=\imgwidth, keepaspectratio]{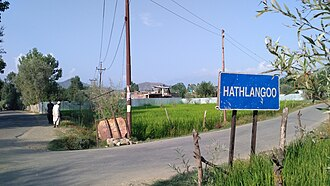}} & \resultdist{147.86} \\
    \cline{1-5} \cline{2-5}
    \resulttxt{dbr:Welland\_Canal}             & \resulttxt{dbr:Playa\_Mayabeque} & \padfig{\includegraphics[height=\imgheight, width=\imgwidth, keepaspectratio]{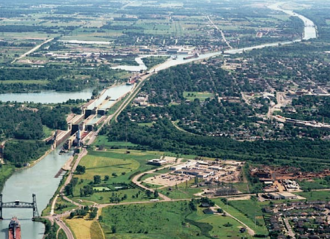}} & \padfig{\includegraphics[height=\imgheight, width=\imgwidth, keepaspectratio]{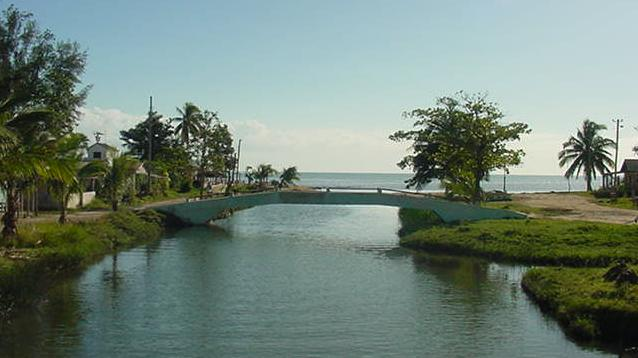}} & \resultdist{147.41} \\
    \cline{1-5} \cline{2-5}
    \resulttxt{dbr:Speikboden\_(South\_Tyrol)} & \resulttxt{dbr:Playa\_Mayabeque} & \padfig{\includegraphics[height=\imgheight, width=\imgwidth, keepaspectratio]{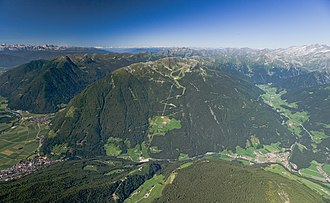}} & \padfig{\includegraphics[height=\imgheight, width=\imgwidth, keepaspectratio]{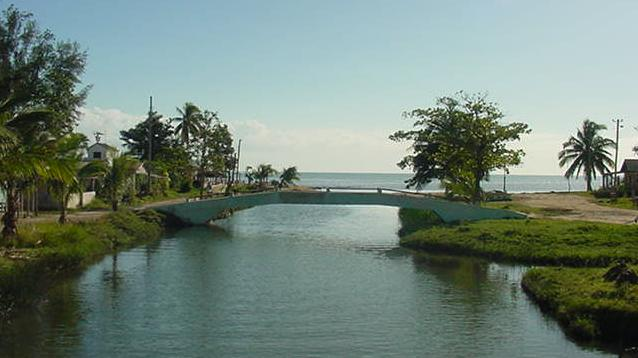}} & \resultdist{143.01} \\
    \cline{1-5} \cline{2-5}
    \resulttxt{dbr:Rim\_(crater)}              & \resulttxt{dbr:Playa\_Mayabeque} & \padfig{\includegraphics[height=\imgheight, width=\imgwidth, keepaspectratio]{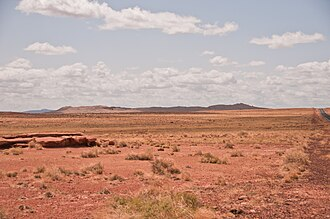}} & \padfig{\includegraphics[height=\imgheight, width=\imgwidth, keepaspectratio]{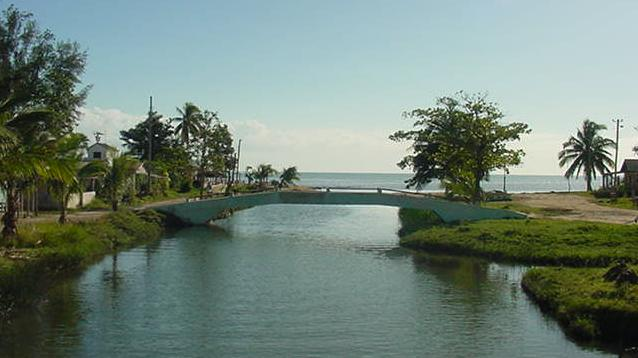}} & \resultdist{140.21} \\
    \cline{1-5} \cline{2-5}
    \resulttxt{dbr:Wittstrauch}                & \resulttxt{dbr:Hathlangoo}       & \padfig{\includegraphics[height=\imgheight, width=\imgwidth, keepaspectratio]{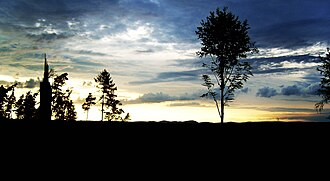}} & \padfig{\includegraphics[height=\imgheight, width=\imgwidth, keepaspectratio]{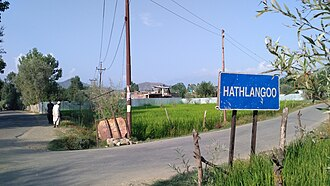}} & \resultdist{139.26} \\
    \cline{1-5} \cline{2-5}
    \tdots                                     & \tdots                           & \vspace{0.5em}                                                                                                                       &                                                                                                                                       &                     \\

    \cline{1-5} \cline{2-5}
    \bottomrule
\end{tabular}

\end{table*}
\tikzexternalenable

\paragraph{Multimodal querying within DBpedia.}

Multimodal querying is one of the key use cases for a real-world large \gls{kg}, enabled by our solution. Our optimisations enable the execution of semantic joins on DBpedia.
Furthermore, we illustrate the outcomes of the join query and show the images used to join in~\Cref{tab:dbpedia_results_hard}. The query returns faithful results based on the textual input and the join condition. The quality of these, however, also strongly depends on the power of the vision model used to encode the images and query, which we do not specifically evaluate in the benchmark campaign, as it is outside our scope.

Nevertheless, the semantic multimodal join returns pairs of scenic villages and remote places. This pairing is reasonable, since they both share open, green features that the image model could register. This complex semantic join operation, combining both graph operations to limit the type to villages and natural places, along with the quadratic lookup, yields a composite query statement. On account of our reduction of the quadratic lookup time to $\bigo{\nt \sqrt{\nt}}$, we can run such queries for the first time.

\section{Conclusion and Future Work}
\label{sec:disc}

This paper proposes integration and advances in dense vector search over \glspl{kg}, prompted by the need for multimodal and dense vector search in emerging \gls{rag} and search pipelines, which currently still rely on separate or unoptimised systems.

To reduce query times for these systems, we introduce three optimisation potentials over existing implementations: 
\begin{enumerate}[label=\alph*.]
    \item the system utilizes fast libraries and builds upon a state-of-the-art \gls{kg} triple store, QLever,
    \item more notably, we shift the parsing of the tensors to the ingestion step of the store, reducing query times significantly.
    \item we offer a way to directly use \gls{aknn} approaches within the engine by offering a virtual, configurable \texttt{SERVICE}, arriving at speedups across datasets, enabling previously infeasible queries.
\end{enumerate}

The improvements are integrated into our \systemname, gaining efficiency improvements by orders of magnitude over the existing systems, thereby enabling queries on large \glspl{kg}. The claims for \systemname are demonstrated on two proposed extended benchmark datasets,
\begin{enumerate}[label=\roman*.]
    \item \bsbmname with text embeddings to demonstrate the scalability of dense vector querying within triple stores,
    
    \item \dbpedianame extended by multimodal image embeddings, showing the real-world impact of optimising these operations; and to demonstrate queries using vector-based joins that could not run in reasonable time spans on prior systems.
\end{enumerate}

Our evaluation compares a range of existing implementations for tensor operations within \acrshort{rdf}, from a two-stage pipeline emulating possible integrations between systems, over an integrated system (RDFTensor), to the \gls{pg} system Neo4j.
We outperform all existing systems using our various optimisation strategies, while also revealing their limitations in applicability. The most notable constraints include the limit on the number of vectors that can be transferred between systems, and the strong limitation of Neo4j's query language to index queries, which only allow very specific constructs. Furthermore, the related system's internal weaknesses are illustrated by our sub-timing benchmark, which shows that computation and parsing of the vectors is one of the key culprits and dominates the query time. These insights lead to our optimisation steps, which we also demonstrated to be the keys to the improved performance through these sub-timing ablations.

The evaluation also unearthed limitations of using indexed approaches on real-world data, where the subsection of the graph, and therefore, indexed vectors might be too small to lead to performance gains over a naive scan compared to the retrieval performance tradeoff. The tensor vocabulary, however, consistently improves efficiency, making it a good default choice for querying dense vectors within \glspl{kg}, regardless of size, query complexity, or application.

The inherent time-recall tradeoff of \gls{aknn} within our system is also evaluated, showing results consistent with existing literature, suggesting that each use case has to weigh the tradeoff between query time and retrieval performance, adjusting parameters accordingly; and for large datasets slight memory overheads as well. Our system, notwithstanding, offers better query timings and resource usage over the existing baseline for either choice.

\paragraph{Limitations and future work}

While the optimisations strive to improve vector similarity search, the inclusion and benchmarking of aggregate functions could be improved to enable even more exotic query types. These, however, are of only secondary interest for applications that combine vector search with \glspl{kg}, the dominant application driving our work.
Further variation of queries used for benchmarking the systems could be extended to include multiple joins across both \gls{kg} structure and semantic similarity, as already noted in the evaluation description. Increasing the query complexity would, however, lead to even less comparability, while having expected complexity bounds.

The speedups and efficiency gains, however, open up new application possibilities within large \glspl{kg}, which can now be queried for a wide range of multimodal data, with cross-encoders already available, be it text, images, sound, or genes~\autocite{Abootorabi2025AskInAnyModality,Bai2026SepRelationAwareKGLearning}.
Further work could also drive novel downstream applications, which will need to integrate with \glspl{lm} to provide fruitful, real-world results. To show applications beyond the current setting with manually written queries and real-world efficiency and, possibly, accuracy gains in benchmarks on \gls{rag} or \gls{qa}, novel integrations will have to be built and tested.
Visual query editors  could now also be extended and tested with the new cross-modal query possibilities, enriching such interfaces with novel interaction modalities. Touching on the \acrshort{rdf} embeddings~\autocite{Ristoski2019RDF2VecRDFGraphEmbeddings}, one could also integrate them into the \acrshort{rdf} \gls{kg}, enabling \emph{integrated \acrshort{rdf} embedding search} over the graph using our optimisation, ending up with a sped up \enquote{Matroschka representations} of the graph.

\printbibliography

\end{document}